\documentclass[twocolumn,showpacs,amsmath,amssymb]{revtex4}
\usepackage{graphicx}
\usepackage{dcolumn}

\begin{document}

\title{Quantum walk approach to search on fractal structures}

\author{E. Agliari}
\email{elena.agliari@fis.unipr.it}
\affiliation{Dipartimento di Fisica, Universit\`{a} degli Studi di Parma,
viale Usberti 7/A, 43100 Parma, Italy}
\affiliation{Theoretische Polymerphysik, Universit\"{a}t Freiburg,
Hermann-Herder-Str. 3, D-79104 Freiburg, Germany}
\author{A. Blumen}%
\affiliation{Theoretische Polymerphysik, Universit\"{a}t Freiburg,
Hermann-Herder-Str. 3, D-79104 Freiburg, Germany}
\author{O. M\"{u}lken}%
\affiliation{Theoretische Polymerphysik, Universit\"{a}t Freiburg,
Hermann-Herder-Str. 3, D-79104 Freiburg, Germany}

\begin{abstract}
We study continuous-time quantum walks mimicking the quantum search
based on Grover's procedure. This allows us to consider structures, that is,
databases,
with arbitrary topological arrangements of their entries. 
We show that the topological structure of the
database plays a crucial role by
analyzing, both analytically and numerically, the transition from the ground to the first excited state of the
Hamiltonian associated with different (fractal) structures. Additionally,
we use the probability of successfully finding a specific target as another
indicator of the importance of the topological structure.
\end{abstract}

\maketitle

\section{Introduction} \label{sec:intro}
In the past two decades quantum computation has attracted growing interest,
encouraged by the development of tools for the manipulation of single quantum
objects as well as by several remarkable theoretical findings \cite{chuang}.
Different systems have been proposed as candidates for quantum computing; they
are based, for instance, on cavity-laser atoms, Bose-Einstein condensates, or NMR techniques (see e.g. \cite{daems,muller,holthaus}). At the same time, a
number of quantum algorithms have been designed and some have been shown to be
even exponentially faster than their best classical counterparts \cite{shor}. In
particular, quantum search algorithms, although able to achieve ``only'' a
polynomial speedup, have been proved to be very promising and of widespread use
in quantum computation \cite{williams,childs3}.

One of the best known quantum search algorithms is attributed to Grover \cite{grover}.
The algorithm can find a target within an unsorted database made up of $N$ items
using $O(\sqrt{N})$ queries. Due to its broad range of applications and its ability to be effectively used as a subroutine \cite{williams,roland2},
Grover's algorithm has been thoroughly investigated and a number of different
implementations have been proposed (see e.g.
\cite{AA,farhi,roland,childs,childs2,daems,dowling,williams,tulsi}). Recently,
implementations based on continuous-time \cite{farhi,childs} as well as on
discrete-time quantum walks \cite{AKR,SKW,tulsi} have been introduced: Given
that the application of random walks in classical algorithms provided
significant advantages for approximations and optimization, one is strongly
motivated to study quantum walks as algorithmic tools. At the current stage
the aim is not only to achieve similar computational improvements, but also to
understand the capabilities of quantum computations. 

Here we focus on the
approach pioneered by Farhi and Gutmann \cite{farhi} and further developed by
Childs and Goldstone \cite{childs}, based on continuous-time quantum walks
(CTQWs). This implementation provides several important advantages: the algorithm does not
need auxiliary storage space and it makes it possible to take into account the geometrical
arrangement of the database (the latter being either a physical position space
or an efficiently encoded Hilbert space \cite{tulsi}). Indeed, while previous
studies just considered the cases of the translationally invariant $\mathbb{Z}^d$
(named $d$-dimensional periodic lattices by Childs and Goldstone \cite{childs}
and called hypercubic lattices in statistical physics, the term which we adopt here) and of complete graphs
\cite{farhi1}, here we extend the investigations to the case of generic
structures and analyze how geometrical parameters [e.g., (fractal)
dimension or (average) coordination number; see later in this article] 
affect the 
dynamics of the CTQW. In \cite{childs} it was shown
that quantum searches based on CTQWs recover the optimal quadratic speedup on
complete graphs and on high-dimensional hypercubic lattices (with dimension
$d>4$), while for low-dimensional ($d<4$) lattices CTQWs can not
outperform their classical counterpart. Hence, it may appear that $d=4$ works as a
``critical dimension'', separating highly performing structures from poorly
performing ones. We will show, both analytically and numerically, that the dimension of the substrate is
not sufficient for getting a sharp transition from the ground to the first excited state; we also take into account the success
probability $\pi_{w,s}(t)$, that is, the probability of finding the quantum
walker at the target site $w$ at time $t$, given as initial state the
equally weighted superposition $s$: An efficient quantum walk gives rise
to a success probability close to $1$ already at very small times $t$.  In
particular, we will take into account several kinds of structures:
translationally invariant structures, such as $d$-dimensional hypercubic
lattices, complete graphs, fractals with low fractal dimension like dual
Sierpinski gaskets and T-fractals, hierarchical structures as
Cayley trees, and structures with (fractal) dimensions larger than four,
such as Cartesian products between Euclidean lattices and dual Sierpinski
gaskets (for the precise definitions of these structures and of the fractal
dimension considered here, see later in this article). In this way we are able to show
that for translationally invariant
structures with high dimensions, $\pi_{w,s}(t)$ displays sharp peaks,
while for fractals or low-dimensional structures
the peaks are low and broad so that the quantum walk is not particularly
effective in the sense that there exists only a low probability
$\pi_{w,s}(t)$ for any $t$. Moreover, for any structure, we evidence
interference phenomena which give rise to a non-monotonic time dependence
of the $\pi_{w,s}(t)$;
such effects can be significant and must be properly taken into account
when considering applications.

Interestingly, the CTQW Hamiltonian used in the quantum search also describes,
in solid-state physics, the dynamics of a tight-binding particle in the presence
of static, substitutional impurities. In this context our results show that in
regular, highly-connected geometries (such as the high-dimensional tori)
the probability of finding the moving particle at the impurity site is
(quasi-) periodic in time and that the localization can be very effective.

Our article is structured as follows. In Sec.~\ref{sec:quantum_search} we first
review basic principles concerning Grover's search and we explain how it can
be implemented by means of CTQWs.  In Sec.~\ref{ssec:inho_graphs} we describe
the structures used as substrate for the CTQW. Then, in
Sec.~\ref{sec:phase_transition} we present several analytical results, later
corroborated and deepened in Sec.~\ref{sec:numerics}, where our numerical
results are shown. In Sec.~\ref{sec:success} we focus on the success
probability. Finally, Sec.~\ref{sec:conclusions} contains our conclusions and
discussion. In Appendix \ref{appe1}, Appendix \ref{appe2} and Appendix
\ref{appe3} we report the details of our analytical calculations.

\section{Quantum walks and Grover's search} \label{sec:quantum_search} 

Grover's search algorithm \cite{grover} is meant to solve the unsorted search
problem under the assumption that there exists a computational oracle
working as a black-box function able to decide whether a candidate
solution is the true solution.  Hence, the oracle knows which is the
target among the $N$ entries.  The task is to find a target $w$ using the
fewest calls to the oracle. While the classical algorithm requires
exhaustive searches implying $O(N)$ queries, Grover's algorithm is able to
find $w$ using $O(\sqrt{N})$ queries, giving rise to a quadratic speed up
\cite{williams,bennett}.

A short outline of the idea behind Grover's algorithm in the presence of a
single marked target is as follows: First of all, one associates each of
the $N$ index integers with a unique orthonormal vector $|x \rangle = |1
\rangle, |2 \rangle, ..., |N \rangle$ in an $N$-dimensional Hilbert space.
Then, one chooses as initial state 
\begin{equation}\label{eq:initial} 
|s \rangle =
\frac{1}{\sqrt{N}} \sum_{x=1} ^{N} |x \rangle, 
\end{equation} 
which is delocalized over the entire set of states  $|x \rangle$ with
equal weights at every site $x$.  This is the least biased initialization
one can arrange, given the available information (since each of the $N$
nodes is in principle equally likely to be the target index, the initial
state is prepared as an equally weighted superposition of all $N$
indices).

Now, to perform the search, one needs to make the state $|s \rangle$ evolve into
a state that has almost all the amplitude concentrated in just the $| w \rangle$
component. In such a way a single final measurement will find the system in the
state $|w \rangle$, hence revealing the identity of the target index.  More
precisely, we need an evolution operator whose repeated application makes the
amplitude of $| w \rangle$ grow with the number of iterations \cite{williams}.
This task was originally accomplished within the standard paradigm for quantum
computation \cite{grover}, namely using a discrete sequence of unitary logic
gates, while in the last years several different implementations have been
introduced in which the state of the quantum register evolves continuously under
the influence of a driving Hamiltonian \cite{farhi1,farhi2,roland}. In
particular, here we follow the approach developed by Childs and Goldstone
\cite{childs} which relies on CTQW \cite{farhi}. As
already mentioned, due to their versatility, CTQWs allow to model any discrete
database. In fact, a generic discretizable database can be represented by a
graph $\mathcal{G}=\{V,E\}$ made up of a set of nodes $V=\{1,2,...,N\}$, each
corresponding to a different item, and of a set of links $E$ joining nodes
pairwise in such a way that the topology of the graph mirrors the arrangement of
the database. The graph $\mathcal{G}$ can be algebraically described by the
adjacency matrix $\mathbf{A}$ whose entry $A_{ij}$ equals one if nodes $i$ and $j$ are
connected, otherwise it is zero (also for the diagonal elements). From $\mathbf{A}$ one can directly calculate
the degree matrix $\mathbf{Z}$, which is a diagonal matrix with elements
$Z_{ij}=z_i \delta_{ij}$, where $z_i = \sum_{j \in N} A_{ij}$ is the
coordination number (or degree) of the $i$-th node, 
that is, the number of its nearest neighbors.

CTQWs on a graph are defined by the Laplacian
matrix $\mathbf{L}=\mathbf{Z}-\mathbf{A}$ 
and obey the following Schr\"{o}dinger equation for the transition
amplitude $\alpha_{k,j}(t)$ from state $| j \rangle$ to state $| k
\rangle$ \cite{kempe}: 
\begin{equation}\label{eq:schrodinger}
\frac{d}{dt} \alpha_{k,j}(t)=-i \sum_{l=1}^{N} H_{kl} \alpha_{l,j}(t),
\end{equation} 
where the Hamiltonian is just given by $\mathbf{H}=\mathbf{L}$, therefore, the time is dimensionless and given in units of the coupling elements $H_{kl}$.
The formal solution for $\alpha_{k,j}(t)$ can be written as
\begin{equation}\label{eq:formal_solution} 
\alpha_{k,j}(t)=\langle k |
\exp(-i\mathbf{H}t)   | j \rangle, 
\end{equation} 
and its magnitude squared provides the quantum mechanical transition
probability $\pi_{k,j}(t) \equiv |\alpha_{k,j}(t)|^2$. Notice that
$\mathbf{L}$ is symmetric and non-negative definite; its ground state,
corresponding to eigenvalue $0$, is given by $| s \rangle$.

Due to the unitary time-evolution generated by $\mathbf{H}$, CTQWs are
symmetric under time-inversion, which precludes $\pi_{k,j}(t)$ from
attaining equipartition at long times. This is different from the behavior
of classical continuous-time random walks. Moreover, CTQWs keep memory of
the initial conditions, as exemplified by the occurrence of (quasi-)
revivals \cite{revival,revival2}.

Now, the ``oracle Hamiltonian'' is given by 
\begin{equation}\label{eq:oracle}
\mathbf{H}_w=-|w \rangle \langle w|, 
\end{equation} 
whose ground state, with energy $-1$, is just the marked item $|w
\rangle$; all other states have energy zero \cite{farhi1}. Then, the
Hamiltonian $\mathbf{H}$ governing the time evolution of the quantum walk
is 
\begin{equation}\label{eq:hamiltonian}
\mathbf{H} = \gamma \mathbf{L} + \mathbf{H}_w, 
\end{equation} 
where $\gamma$ is a tunable parameter with units of inverse time, hence dimensionless here.

The \textit{success probability} $\pi_{w,s}^{\gamma}(t)$ is defined as
(see also \cite{childs}) 
\begin{equation}\label{eq:success_prob} 
\pi_{w,s}^{\gamma}(t)
\equiv |\langle w | \exp(-i \mathbf{H} t) | s \rangle|^2,  
\end{equation} 
i.e., as the transtion probability to be at time $t$ at the target $w$
when starting in the delocalized state $|s\rangle$.

Here we study its dependence on time $t$ and on the parameter $\gamma$ and
we especially aim to evidence the existence of an optimization parameter
$\gamma_\mathrm{max}$, possibly depending on time, which maximizes
$\pi_{w,s}^{\gamma}(t)$. Notice that, having fixed $\gamma$, the time $t$
can be measured in terms of the number of queries of the discrete Grover
oracle \cite{childs2}.

\section{Fractal and hierarchical structures}\label{ssec:inho_graphs} 

\begin{figure} \includegraphics[width=55mm,height=48mm]{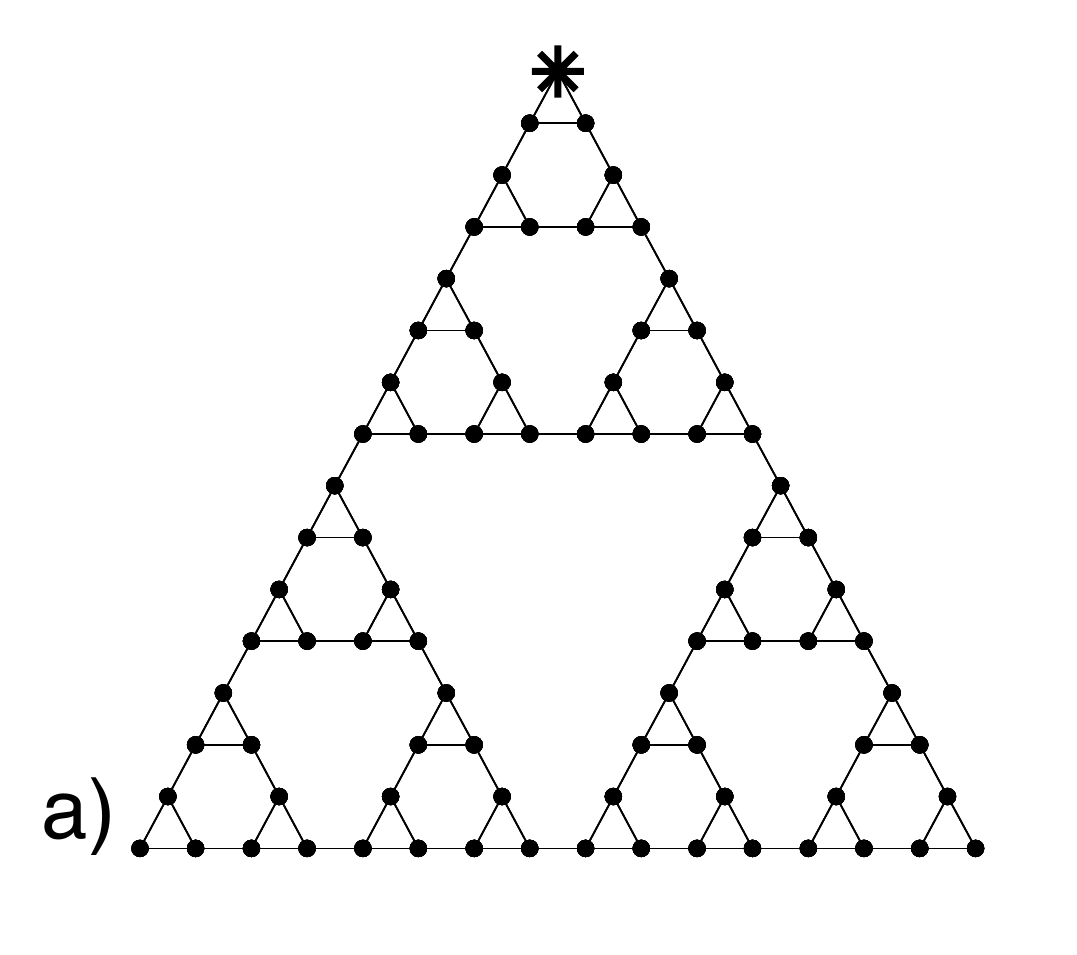}
\includegraphics[width=55mm,height=48mm]{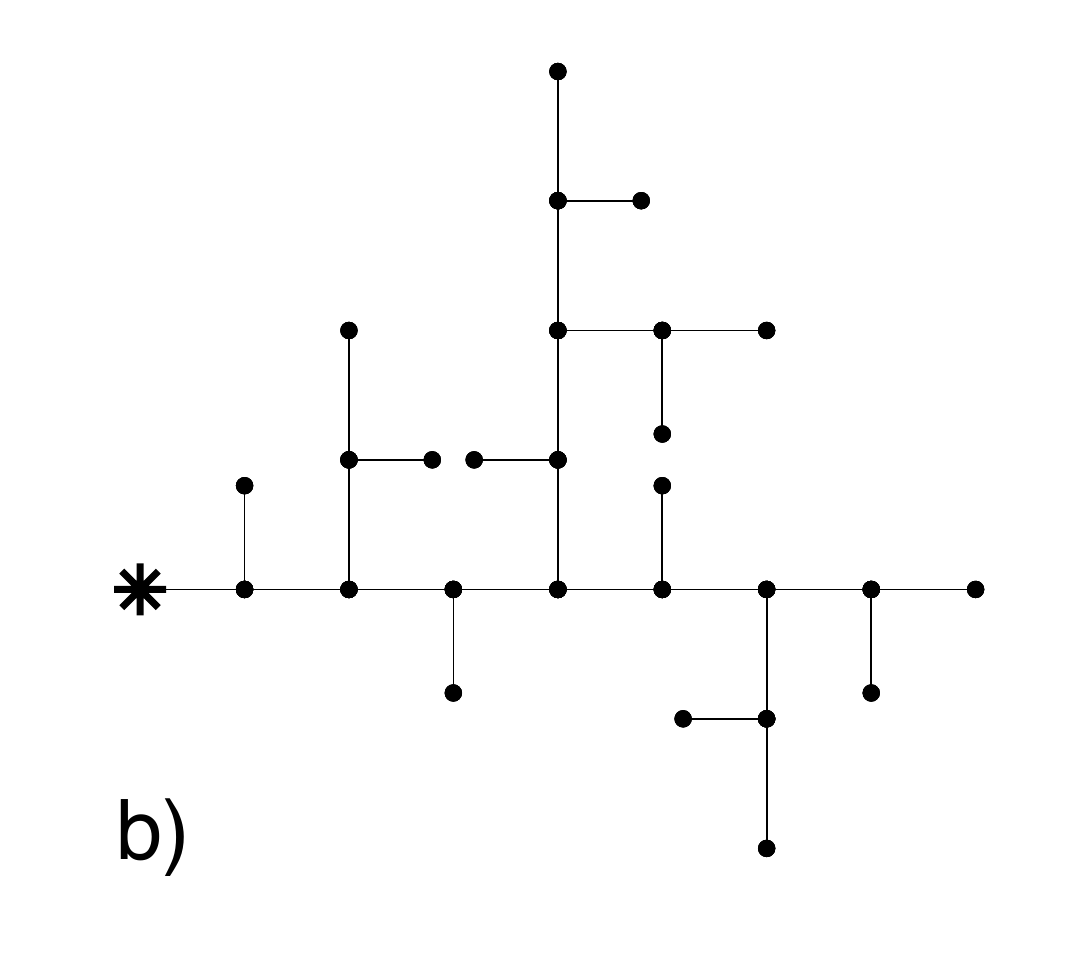}
\includegraphics[width=55mm,height=48mm]{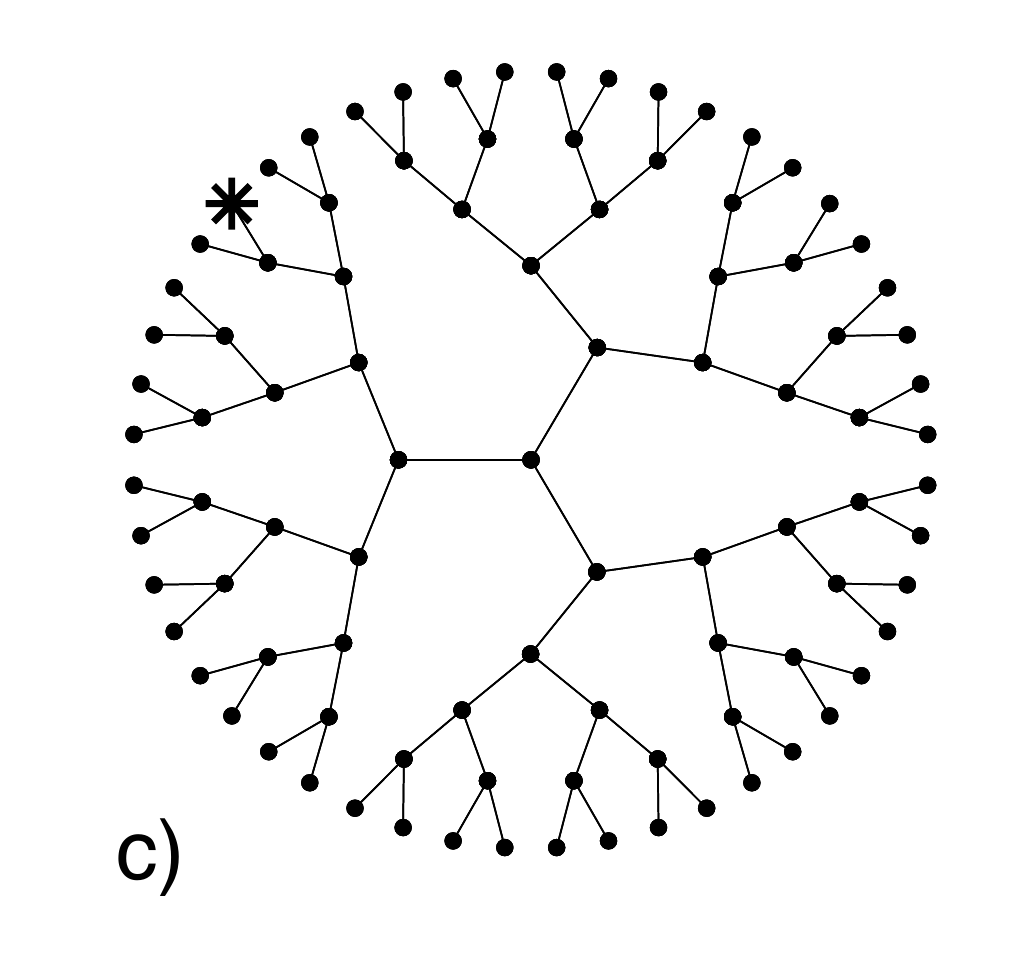}
\caption{\label{fig:frattali}Examples of fractal structures considered here; in
general the star indicates the position of the trap. Panel $a$: Dual Sierpinski
gasket of generation $4$ and volume $N=3^g=81$; due to symmetry, the
three corners are equivalent. Panel $b$: T-fractal of generation $3$ and volume
$N=3^g+1=28$; due to the symmetry all outmost sites (at distance $2^{g-1}$ from
the central node) are equivalent. Panel $c$: Cayley tree of generation $5$ and
volume $N=3 \times 2^{g}-2=94$; all of the $N/2+1$ outmost sites are
equivalent.} \end{figure}

Before proceeding, it is worth introducing the geometrical structures on
which we are focusing, first the dual Sierpinski gasket (DSG), the
T-fractal (TF), and the Cayley tree (CT); examples of these structures are
shown in Fig.~\ref{fig:frattali}a, Fig.~\ref{fig:frattali}b, and
Fig.~\ref{fig:frattali}c, respectively.  Notice that these structures
differ significantly from those analyzed previously: While hypercubic
lattices with periodic boundary conditions (toroids) 
are translationally invariant, the aforementioned structures are not.

The DSG, TF and CT can be built iteratively; at the $g$-th iteration we
have the fractal of generation $g$ (see e.g. \cite{ABM2008,redner,cosenza}). The
DSG and the TF are examples of exactly decimable fractals for which the fractal
dimension, $d_f$,
and the spectral dimension, $\tilde{d}$, are exactly known.  
While the fractal dimension relates the number of nodes inside a
sphere to the radius of the sphere \cite{fractals}, the
spectral dimension is obtained from the scaling of the eigenmodes of a given
structure (phonons for lattices, fractons for fractal substrates)
\cite{sdim}, most simply seen in the probability of a random
walker to be (still or again) at the original site \cite{aodim}.

Here, we have namely $d_f = \ln 3 / \ln 2 \approx 1.585$ and $\tilde{d}= 2
\ln 3 / \ln 5 \approx 1.365$ for the DSG, and $d_f = \ln 3 / \ln 2$ and
$\tilde{d}= 2 \ln 3 / \ln 6 \approx 1.226$ for the TF. We recall that for
the usual, translationally invariant lattices the spectral and fractal
dimensions coincide with the Euclidean
dimension $d$, namely $\tilde{d}=d_f=d=1$ for the chain, $\tilde{d}=d_f=d=2$ for the square
lattice and so on. On the other hand, on fractal
structures $\tilde{d}$ often replaces $d$ when dealing with dynamical and
thermodynamical properties \cite{hattori}. Also, for fractals
$\tilde{d}<d_f$ and $d_f$ is smaller
than the Euclidean dimension of the embedding space, that is, 2 for DSG and
TF. The CT is no fractal in the classical sense, since its growth with
increasing generation is exponential. However, the CT is built in a
hierarchical manner, analogous to the TF. We also notice that both, the TF and
the CT, are trees (and are hence devoid of loops) and exhibit a large
number $O(N/2)$ of end nodes.

As we will show in the following, CTQW on
the DSG, the TF and the CT can display a low
probability $\pi_{w,s}^\gamma(t)$ for any $t$ and $\gamma$ when compared
with the case of the translationally invariant lattices, especially when the
dimension of the lattice becomes larger than $4$.
Therefore one can ask whether the probabilities $\pi_{w,s}^{\gamma}(t)$ can be
improved by adopting (fractal or hierarchical) substrates which display a
large spectral dimension and a large coordination number.

For instance, we can build up such structures by combining the DSG and Euclidean lattices by means of Cartesian products.  In general,
the Cartesian product of two graphs $\mathcal{G}_1=\{V_1,E_1\}$ and
$\mathcal{G}_2=\{V_2,E_2\}$ is a graph $\mathcal{G} = \mathcal{G}_1 \times
\mathcal{G}_2$ with the vertex set $V_1 \times V_2$, and such that two nodes
$(x_1 x_2, y_1 y_2)$ are adjacent if $(x_1,y_1) \in E_1$ and $x_2 = y_2$, or
$x_1=y_1$ and $(x_2, y_2) \in E_2$. It has been shown \cite{woess} that the
spectral dimension $\tilde{d}$ on the product graph $\mathcal{G}$ is then the sum of
the corresponding dimensions of the two initial graphs $\tilde{d}_1$ and
$\tilde{d}_2$.

We combine in this way the DSG with the chain $L_1$, with the square lattice $L_2$ and
with the cubic lattice $L_3$, to obtain more complex structures displaying spectral
dimensions approximately equal to $2.365$, $3.365$ and $4.365$, respectively.

Finally, it should be underlined that, dealing with such structures, the
location of the target, i.e. the node $w$, also (quantitatively) matters. In the
numerical analysis of Sec.~\ref{sec:numerics} we place the target on a
peripheral site, which means, without loss of generality, the apex for the DSG,
the leftmost site for the TF and an outmost site for the CT (see
Fig.\ref{fig:frattali}). We expect that a peripheral position for the target
site does not correspond to an optimal situation and, since a priori the target
position is unknown, this choice prevents an overestimation of the
probability of finding the target.

\section{Overlaps and transition}\label{sec:overlaps_transitions} 

Let us consider the Hamiltonian of Eq.~(\ref{eq:hamiltonian}) and denote
the corresponding set of eigenstates and eigenvalues by $\{|\psi_k\rangle
\}$ and $\{ E_k  \}$, respectively. Now, for large $\gamma$ the
contribution of $\mathbf{H}_w$ to $\mathbf{H}$ is negligible and the
ground state $|\psi_0\rangle$ is close to $|s\rangle$. On the other hand,
as $\gamma \rightarrow 0$ the ground state is close to $|w\rangle$ since
the weight of $\mathbf{L}$ is small and, from perturbation theory, we
expect that $| s \rangle$ is close to $| \psi_1 \rangle$, i.e. to the
first excited state of $\mathbf{H}$ \cite{childs}.

As pointed out by Childs and Goldstone \cite{childs}, there exists an
intermediate range of $\gamma$ where, for complete graphs and hypercubic
lattices with dimension $d>4$, $|s \rangle$ switches from the ground state $|
\psi_0 \rangle$ to the first excited state $| \psi_1 \rangle$; in the very same
region of $\gamma$ the target state $| w \rangle$ switches from a state with
large overlap with $| \psi_1 \rangle$ to the ground state $| \psi_0 \rangle$.
Therefore, by varying $\gamma$, we can find a particular value for $\gamma$ for which the Hamiltonian $\mathbf{H}$ can evolve the state $|s \rangle$ into a state close to the target state $|w \rangle$.
The existence and the narrowness of such a range for $\gamma$ are crucial for a
large success probability.

\subsection{Analytical results}\label{sec:phase_transition} 

As anticipated, the CTQW under study can yield effective results if the
Hamiltonian $\mathbf{H}$ is able to rotate the state $| s \rangle$ to a
state with a large overlap with $| w \rangle$. For this to occur a first
condition which needs to be fulfilled is that there exists an intermediate range of $\gamma$ over which the state $| s \rangle$ has a substantial overlap with both $| \psi_0 \rangle$ and $| \psi_1 \rangle$. In this subsection the
occurrence of such a condition is investigated analytically for an
arbitrary structure, following arguments similar to those exploited in
\cite{childs} for $d$-dimensional hypercubic lattices; the details of the
calculations are given in Appendixes A and B
while here we just report the main results.

First of all, we define $| \phi_k \rangle$ and $\mathcal{E}(k)$, the
$k$-th eigenstate and eigenvalue of the Laplacian $\mathbf{L}$,
respectively. In the basis of the eigenstates $| \phi_k \rangle$ the
target site $|w \rangle$ can be written as 
\begin{equation} \label{eq:expansion} 
|w \rangle =  \sum_{k=0}^{N-1}
a_k |\phi_k \rangle, 
\end{equation} 
where $a_k \equiv \langle w| \phi_k \rangle$.

As derived in Appendix~\ref{appe1}, once $\xi_j \equiv \sum_{k \neq 0}
|a_k|^2 /[\mathcal{E}(k)]^j$ is set, the overlap of $| s \rangle$ with the
ground state $| \psi_0 \rangle$ or with the first non-degenerate excited
state $| \psi_1 \rangle$ is (depending on $\gamma$) limited from below by
the same bound. We find namely 
\begin{equation} \label{eq:s_0_short} 
1> |\langle s | \psi_0 \rangle
|^2 > 1 - \frac{\xi_2}{N (\gamma - \xi_1 )^2}, 
\end{equation} 
for $\gamma > \xi_1$, and also 
\begin{equation} \label{eq:s_l_short} 
1> |\langle s | \psi_1
\rangle |^2 > 1 - \frac{\xi_2}{N (\xi_1 - \gamma)^2}, 
\end{equation} 
for $\gamma < \xi_1$. 

Now, we define the (size dependent) critical value
$\tilde\gamma$ as that value of $\gamma$ for which
\begin{equation}\label{eq:intersection}
|\langle s | \psi_0\rangle |^2 = |\langle s|\psi_1\rangle|^2
\end{equation}
is satisfied, i.e., the $\gamma$-value for which the projection of state $s$ onto the ground and the first excited state has the same magnitude. This is particularly interesting in the limit $N \to \infty$, when a level crossing from state $| \psi_0 \rangle$ to state $| \psi_1 \rangle$ can occur. In fact, we find that if $\xi_2 /[N(\xi_1 - \gamma)^2]$ converges (at least in the limit $N
\rightarrow \infty$) to $0$ as $\gamma$ approaches $\xi_1$ (from different
sides), then $|\langle s | \psi_0 \rangle |$ and $|\langle s | \psi_1
\rangle |$ both approach $1$ (see Eqs.~(\ref{eq:s_0_short}) and
(\ref{eq:s_l_short})), namely a transition occurs at $\gamma = \xi_1$.  Notice, however, that the
condition for this to occur is in general non-trivial as $\xi_1$ and
$\xi_2$ both depend on $N$. In Appendix~\ref{appe1} we find a sufficient
condition in the Laplacian spectrum and, in particular, in
Appendix~\ref{appe2} we prove that such a condition holds for the DSG for
which $\mathcal{E}(k)$ is exactly known \cite{cosenza,jurjiu,ABM2008}; in
this case we find that
\begin{equation}\label{eq:xi1_asy_short} 
\xi_1 \leq C N^{\alpha + 2/\tilde{d}},
\end{equation} 
and 
\begin{equation}\label{eq:xi2_asy_short} 
\xi_2 \leq C
N^{\alpha + 4/\tilde{d}}, 
\end{equation} 
where $\alpha$ is a parameter
depending on the given network. In general, $-1 \leq \alpha < 0$; for
hypercubic lattices $\alpha=-1$, regardless of their dimension, while for
the DSG we can numerically estimate $\alpha$ as being $\alpha \approx
-0.9$ (see Appendix~\ref{appe1}). Therefore, we expect (for DSG)
$\tilde\gamma$ to be, approximately:
\begin{equation}\label{eq:gammacrit_short} 
\tilde{\gamma} \approx C
N^{\alpha+2/\tilde{d}}.  
\end{equation} 
This result is consistent up to logarithmic corrections with the critical
points found in \cite{childs} for the linear chain and the square lattice,
namely $\tilde{\gamma} \sim N$ and $\tilde{\gamma} \sim \log N$,
respectively.

Finally, we point out that the critical parameter $\tilde{\gamma}$
provides interesting information in the context of quantum adiabatic computation
\cite{roland,santoro}: $\tilde{\gamma}$ represents a threshold below which we
can expect $| w \rangle$ to have a large overlap with the ground state.

\subsection{Numerical results}\label{sec:numerics} 

\begin{figure} \includegraphics[width=80mm,height=70mm]{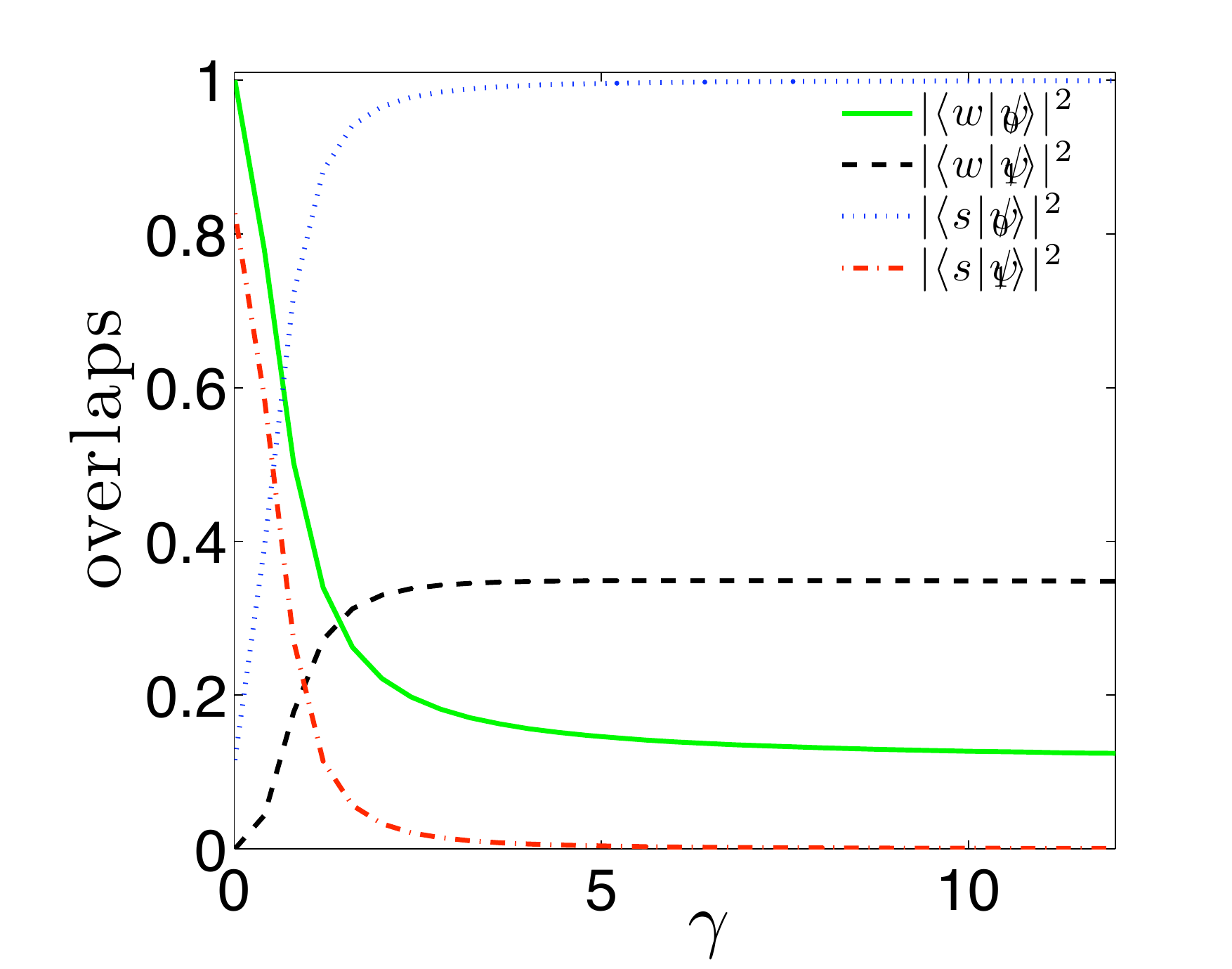}\\
\includegraphics[width=80mm,height=70mm]{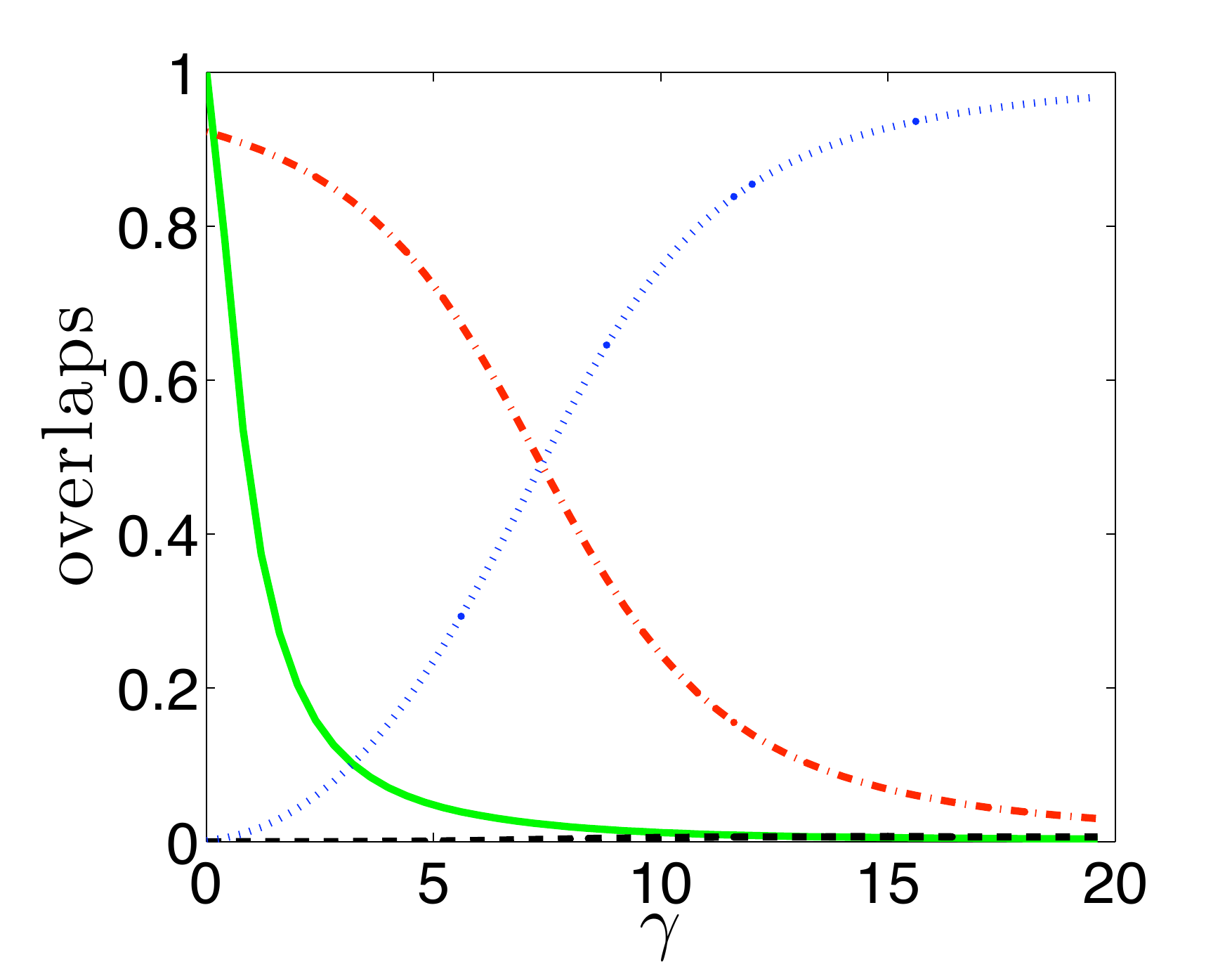}
\caption{\label{fig:DS_ext}(Color online) Overlaps for a DSG of generation $g=3$
(up) and $g=6$ (bottom) as a function of (the dimensionless) $\gamma$, see text for details.} \end{figure}

\begin{figure} \includegraphics[width=80mm,height=70mm]{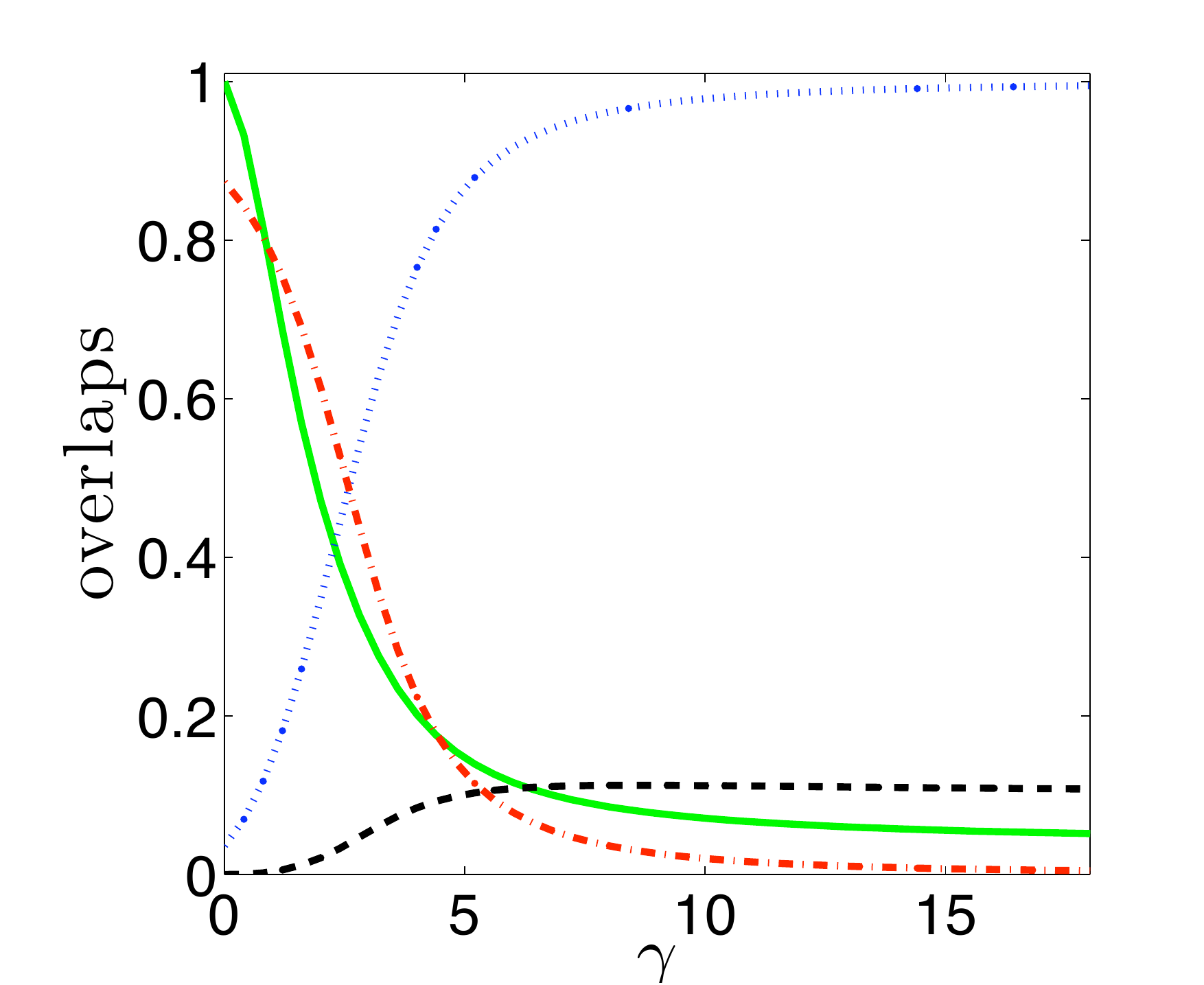}\\
\includegraphics[width=80mm,height=70mm]{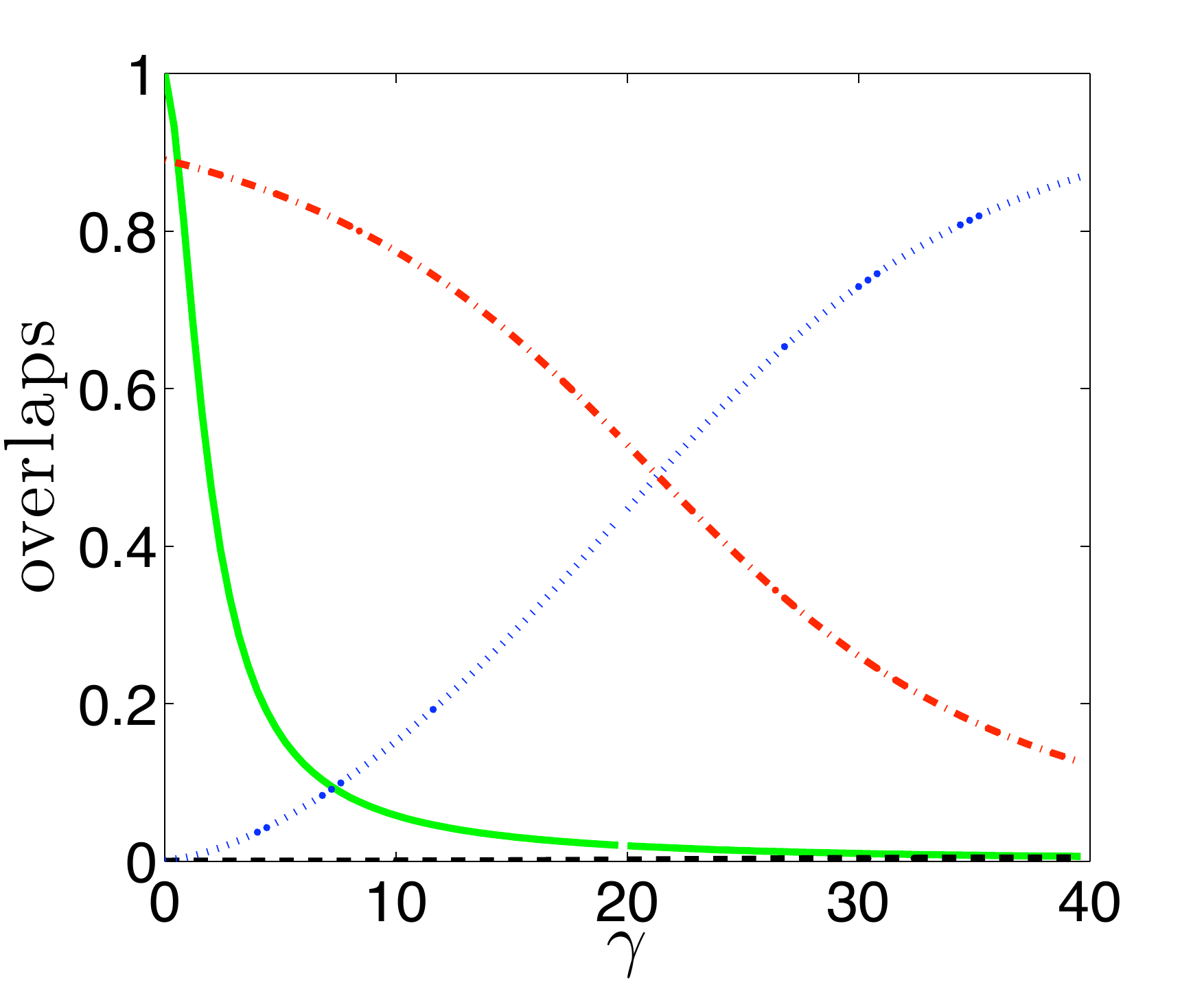}
\caption{\label{fig:T_ext}(Color online) Overlaps for a T-fractal of generation
$g=3$ (up) and $g=6$ (bottom) as a function of $\gamma$. The symbols are as in
Fig.~\ref{fig:DS_ext}.} \end{figure}

\begin{figure} \includegraphics[width=80mm,height=70mm]{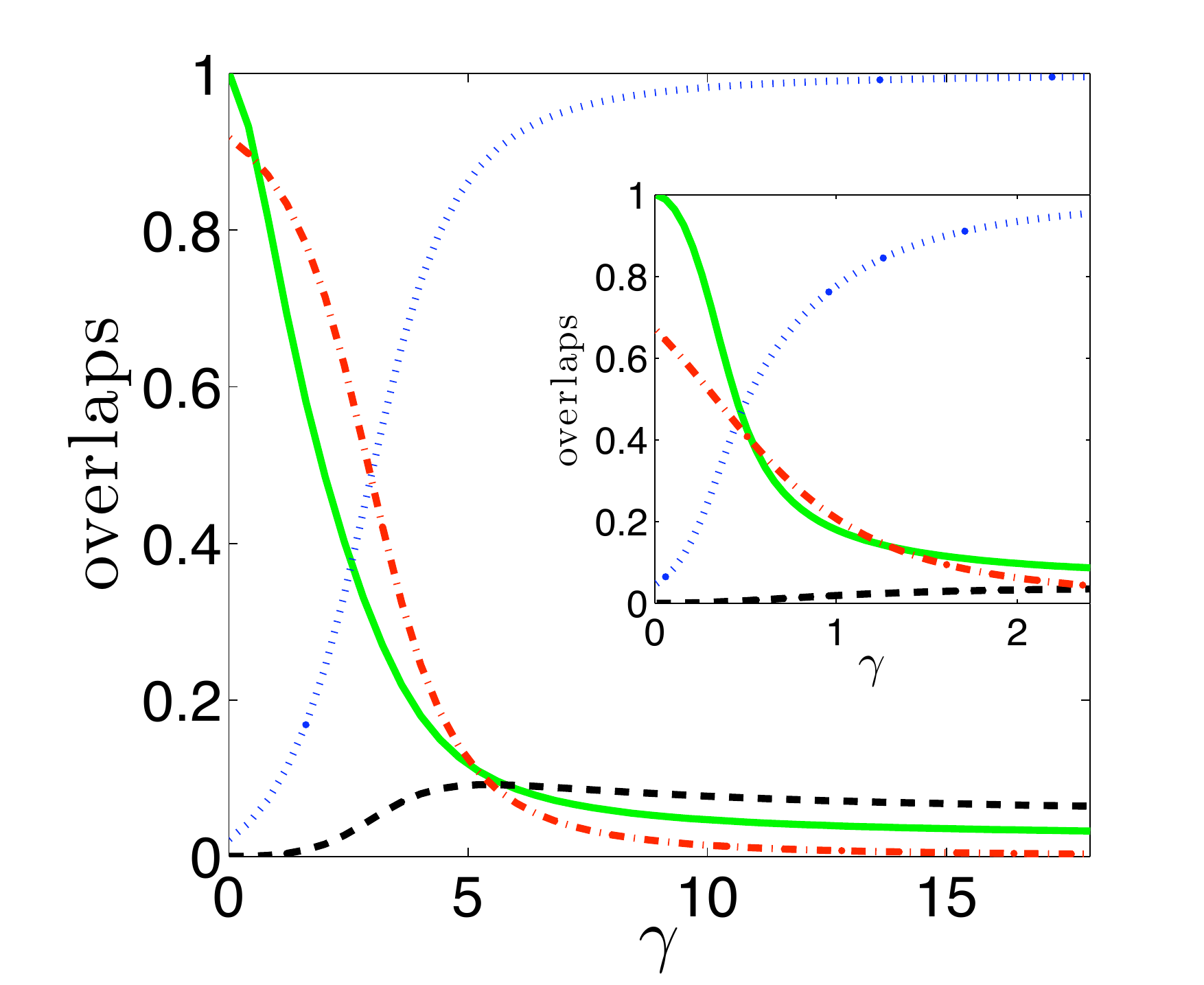}\\
\includegraphics[width=80mm,height=70mm]{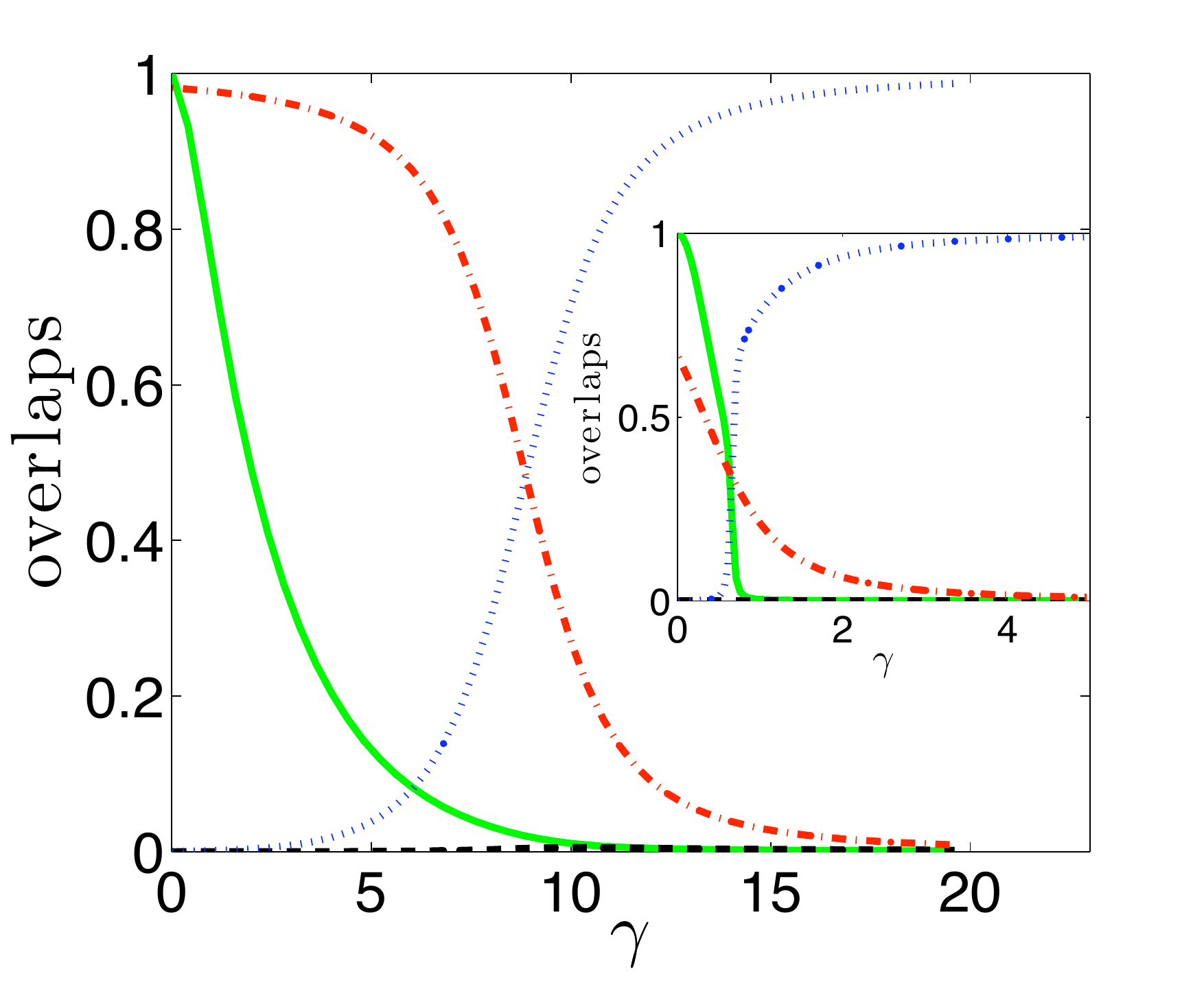}
\caption{\label{fig:CT_ext}(Color online) Overlaps for a CT of generation $g=3$
(up) and $g=10$ (bottom) as a function of $\gamma$. The symbols are as in
Fig.~\ref{fig:DS_ext}.} \end{figure}

We now consider the three examples of low-dimensional inhomogeneous
structures described previously, for which the overlaps of the initial state
$|s \rangle$ and of the target state $|w \rangle$ with $|\psi_1 \rangle$
and $|\psi_0 \rangle$ are shown in Figs.~\ref{fig:DS_ext}-\ref{fig:CT_ext}
for different generations $g$, as a function of the parameter $\gamma$.
These plots evidence that there exists an intermediate range of $\gamma$
where the state $|s \rangle$ changes from having a large overlap with the
first excited state to having a large overlap with the ground state. In
the same region of $\gamma$ the overlap $|\langle w | \psi_1 \rangle|^2$
is significant for structures of small size (top panels in
Figs.~\ref{fig:DS_ext}-\ref{fig:CT_ext}), while it is still very small
when the size is large (bottom panels in
Figs.~\ref{fig:DS_ext}-\ref{fig:CT_ext}). This is vastly distinct from the
situation found for hypercubic lattices \cite{childs} where close to
$\tilde{\gamma}$ the overlap $|\langle w | \psi_1 \rangle|^2$ is
significant.

By comparing the plots obtained for the DSG (Fig.~\ref{fig:DS_ext}),
the TF (Fig.~\ref{fig:T_ext}), and the CT (Fig.~\ref{fig:CT_ext}), we
notice that $\tilde{\gamma}$ depends sensitively on the underlying
topology. In fact, going from DSG to CT and then to TF we notice an
amplification of the transition region, which is for TF most spread out,
requiring relatively large values of $\gamma$ in order for $| s \rangle$
to have a large overlap with the ground state.  Such effects can be
ascribed to the absence of loops and to the existence for TF of a large
number of peripheral sites scattered throughout the whole structure which
give rise to localization effects \cite{ABM2008}.

\begin{figure} \includegraphics[width=80mm,height=70mm]{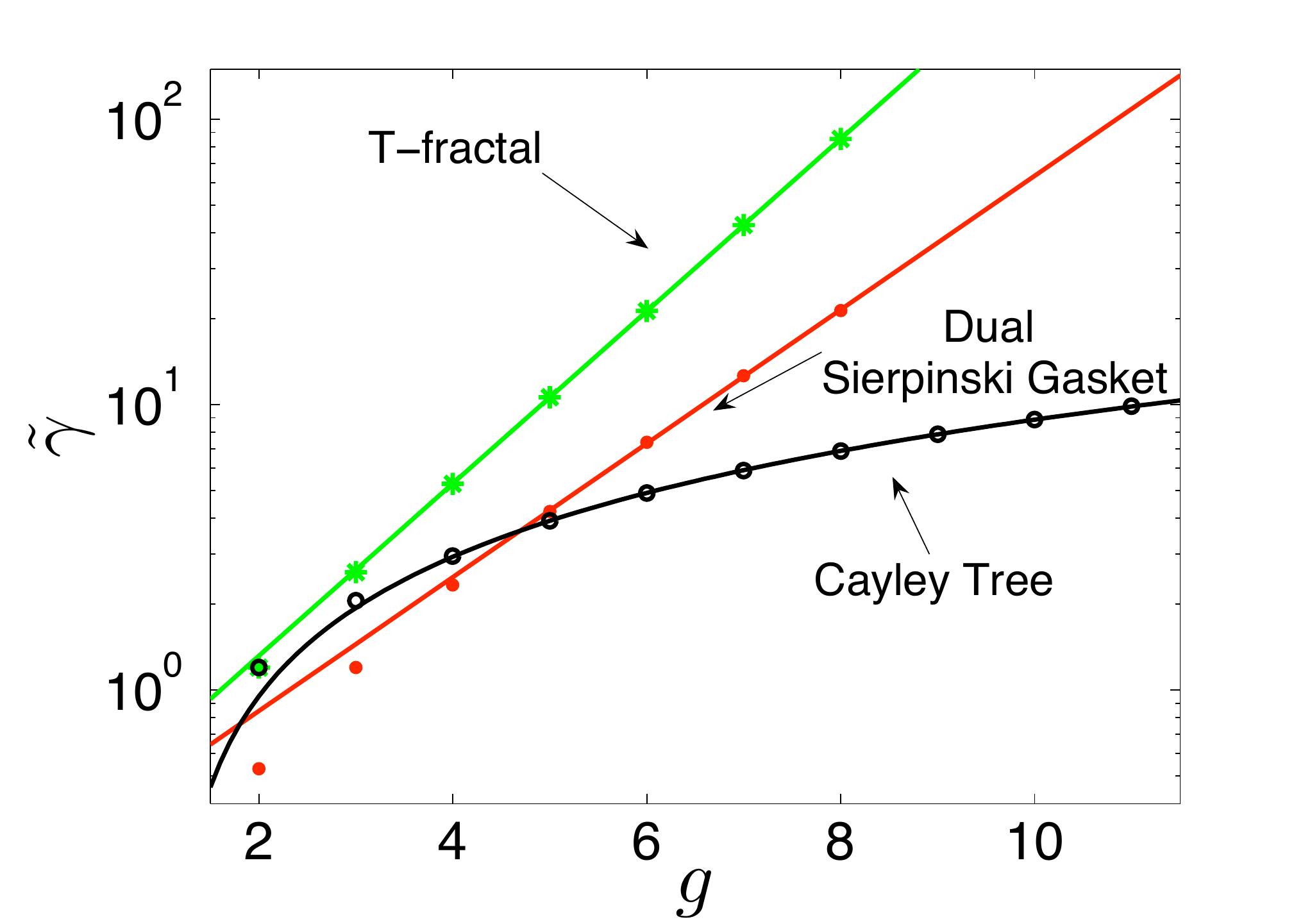}
\caption{\label{fig:gamma_crit}(Color online) Estimate of $\tilde{\gamma}$
for DSG ($\bullet$), TF ($*$) and CT (o) in a semi-logarithmic scale. The
continuous lines represent the best fits.} \end{figure}

We now calculate $\tilde{\gamma}$ according to Eq.~(\ref{eq:intersection})
and for several values of $g$; numerical data and relative best fits are
depicted in Fig.~\ref{fig:gamma_crit}.  For the DSG, numerical points are
best fitted by the function $\tilde{\gamma} \approx 3^{0.55 g} \approx
N^{0.55}$, in very good agreement with the analytical findings. In fact,
according to our analytical investigations, $\tilde{\gamma}$ scales with
the size of the database like $N^{\alpha + 2/\tilde{d}}$ (see
Eq.~(\ref{eq:gammacrit_short})), where, for the DSG, $\alpha + 2/\tilde{d}
= \alpha +  \log 5 / \log 3 \approx 0.57$  ($\alpha$ is taken to equal
$-0.9$).  Let us now consider the case of the TF: From numerical data the
best fit turns out to be $\tilde{\gamma} \approx 3^{0.70 g} \approx
N^{0.70}$.  Interestingly, this result is still in very good agreement
with the analytical approximation of Eq.~(\ref{eq:gammacrit_short}). In fact,
for the TF the exponent gets  $\alpha +  2/\tilde{d} = \alpha +  \log 6 /
\log 3 \approx 0.73$ (where, again, $\alpha$ is taken to equal $-0.9$,
consistently with the estimates given in Appendix \ref{appe1}).  Such a
consistency might suggest that Eq.~(\ref{eq:gammacrit_short}) is valid not only
for the DSG but for any (exactly decimable) fractal with $\tilde{d}<2$.
As for the CT, not being a fractal, Eq.~(\ref{eq:gammacrit_short}) does
not hold. Indeed, we find that the value of $\tilde{\gamma}$ corresponding
to Eq.~(\ref{eq:intersection}) grows linearly with the generation of the
fractal, namely logarithmically with $N$. In Fig.~\ref{fig:gamma_crit}
numerical data are fitted by the curve $\tilde{\gamma} = g -1$ (notice the
semilogarithmic plot).

Apart from this, the plots shown in
Figs.~\ref{fig:DS_ext}-\ref{fig:CT_ext} look rather similar. In
particular, for networks of large sizes $|\langle w| \psi_0 \rangle|^2$
decays with $\gamma$ more rapidly than $|\langle s| \psi_1 \rangle|^2$.
Hence, the range of $\gamma$ over which the transition occurs is wide,
analogously to what happens on low-dimensional hypercubic lattices (see
\cite{childs}). As shown in the next section, this has important effects
on the behaviour of the success probability and suggests that a sharp transition is associated with an effective search algorithm.

In order to sketch the role of the position of a target we show in the inset of
Fig.~\ref{fig:CT_ext} for the CT of $g=10$ the case of a target placed on a nearest neighbor of the central node. We see that the transition region is shifted
towards lower values of $\gamma$. This means that a more central placement of
the target is improving the probability for the CTQW to reach the target. Analogous results were also found for DSG and for TF. Thus, limiting our focus to peripheral nodes will prevent us from overestimating the success probabilities.

We now consider fractal structures exhibiting large spectral dimension; in
particular, we focus on fractals obtained from Cartesian products, such as
DSG $\times$ $L_1$, DSG $\times$ $L_2$ and DSG $\times$ $L_3$, as
introduced in the previous section. Again, we place the target on a
``peripheral site'', i.e. on a minimally connected site; this displays the
coordination number $5$, $7$ and $9$, for DSG $\times$ $L_1$, DSG $\times$
$L_2$ and DSG $\times$ $L_3$, respectively.  We calculate for these
structures the overlaps of the initial state $|s \rangle$ and of the
target state $| w \rangle$ with $|\psi_0 \rangle$ and $|\psi_1 \rangle$;
results for DSG $\times$ $L_2$ ($\tilde{d} = 3.365$) and for DSG $\times$
$L_3$ ($\tilde{d} = 4.365$) are shown in Fig.~$6$.

\begin{figure} \label{fig:cartesian}
\includegraphics[height=70mm]{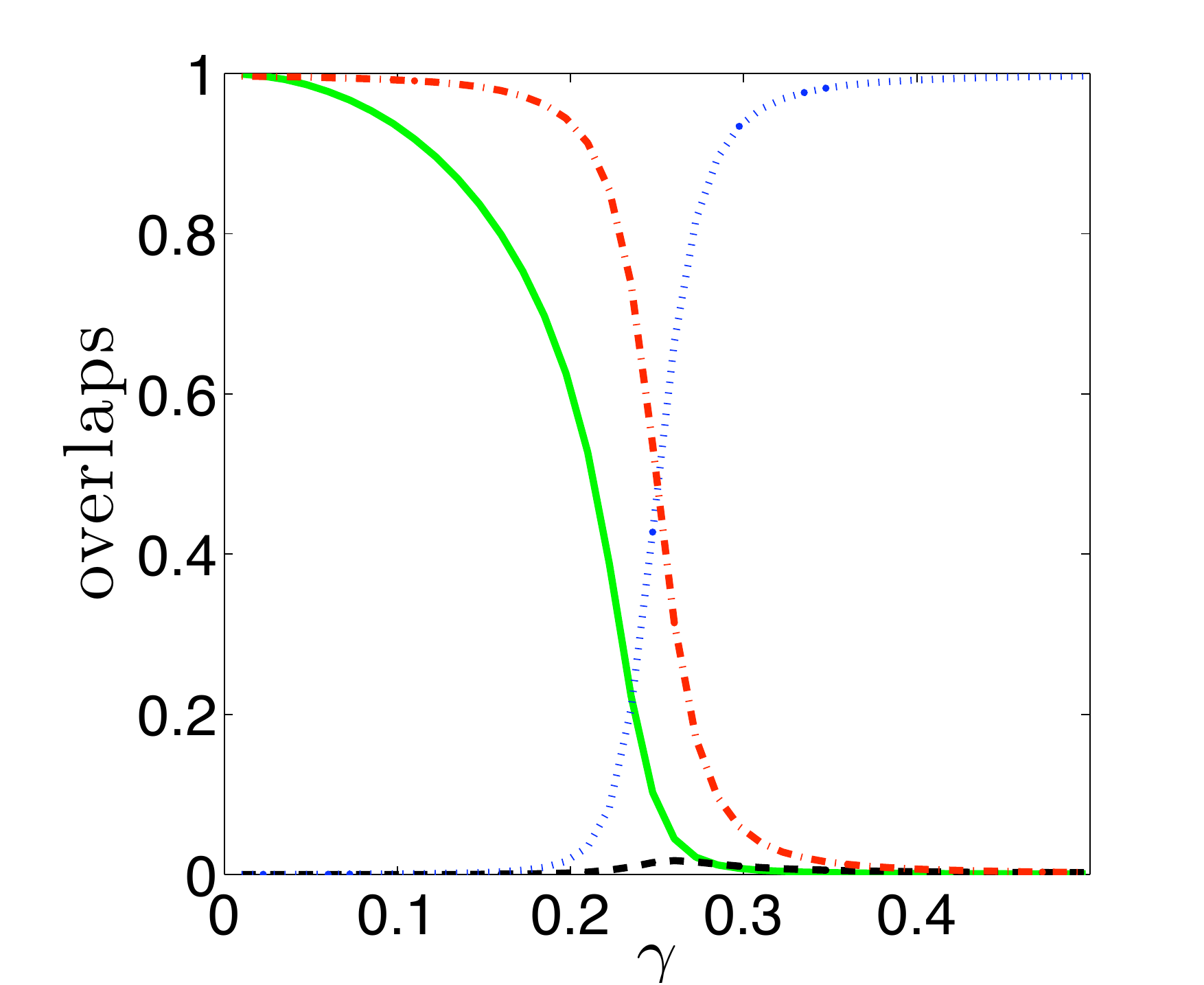}
\includegraphics[height=70mm]{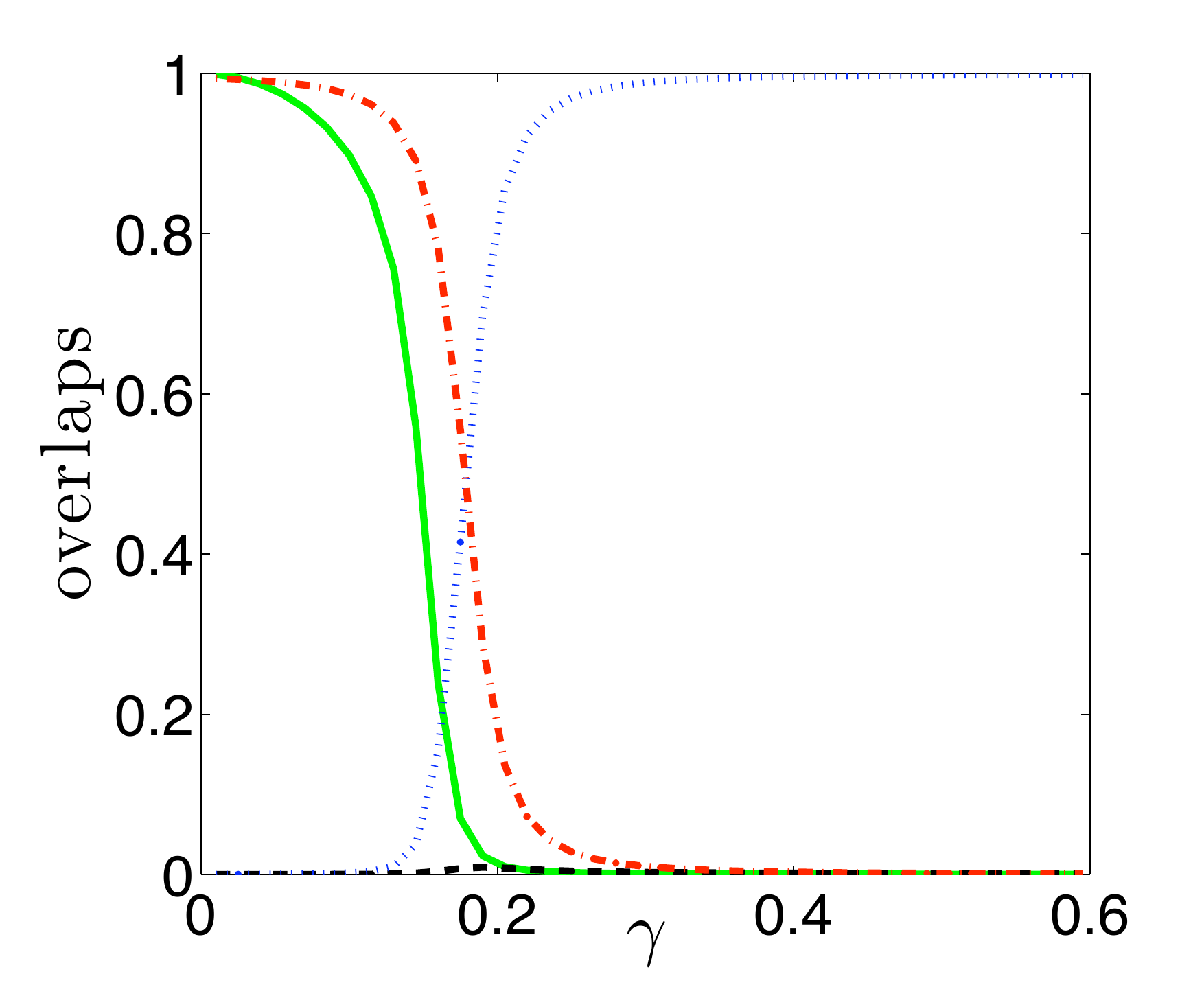} \caption{
(Color online) Upper panel: Overlaps for a DSG $\times$ $L_2$ structure given by
the Cartesian product of a DSG of generation $g=4$ and a square lattice of size
$L=8$; Bottom panel: Overlaps for a DSG $\times$ $L_3$ obtained from a DSG of
generation $g=4$ and a cubic lattice of size $L=4$. The lines are as in
Fig.~\ref{fig:DS_ext}.} \end{figure}

As stressed at the beginning of this section, these plots provide some
information about the sharpness of the transition from
state $| s \rangle$ to the state $|\psi_0 \rangle$ and from state $| w
\rangle$ to the state $|\psi_0 \rangle$. However,
around $\tilde{\gamma}$ also a significantly large overlap $|\langle w |
\psi_1 \rangle|^2$ is required.

Here,  the transitions are still rather smooth although, by increasing
$\tilde{d}$, the region of $\gamma$ values, over which the curves
representing $|\langle w | \psi_0 \rangle|^2$, $|\langle s | \psi_0
\rangle|^2$ and $|\langle s | \psi_1 \rangle|^2$ intersect, is shrinking.
On the other hand, the overlaps between $| w \rangle$ and the first
excited state $| \psi_1 \rangle$ are negligible for all values of
$\gamma$. 

\section{Success Probability} \label{sec:success} 

We now turn to the success probability $\pi_{w,s}^{\gamma}(t)$,
Eq.~(\ref{eq:success_prob}), and we investigate numerically its dependence
on $t$ and on $\gamma$. Because of its dependence on time,
$\pi_{w,s}^{\gamma}(t)$ carries more information than the previously
discussed overlaps.

We first analyze the case of complete graphs and of $d$-dimensional hypercubic
lattices (for which the time dependences of $\pi_{w,s}^{\gamma}(t)$ have already
been determined for special choices of $\gamma$ in Ref. \cite{childs}) before
turning to the DSG, the TF and the CT.

\begin{figure} \includegraphics[width=85mm,height=70mm]{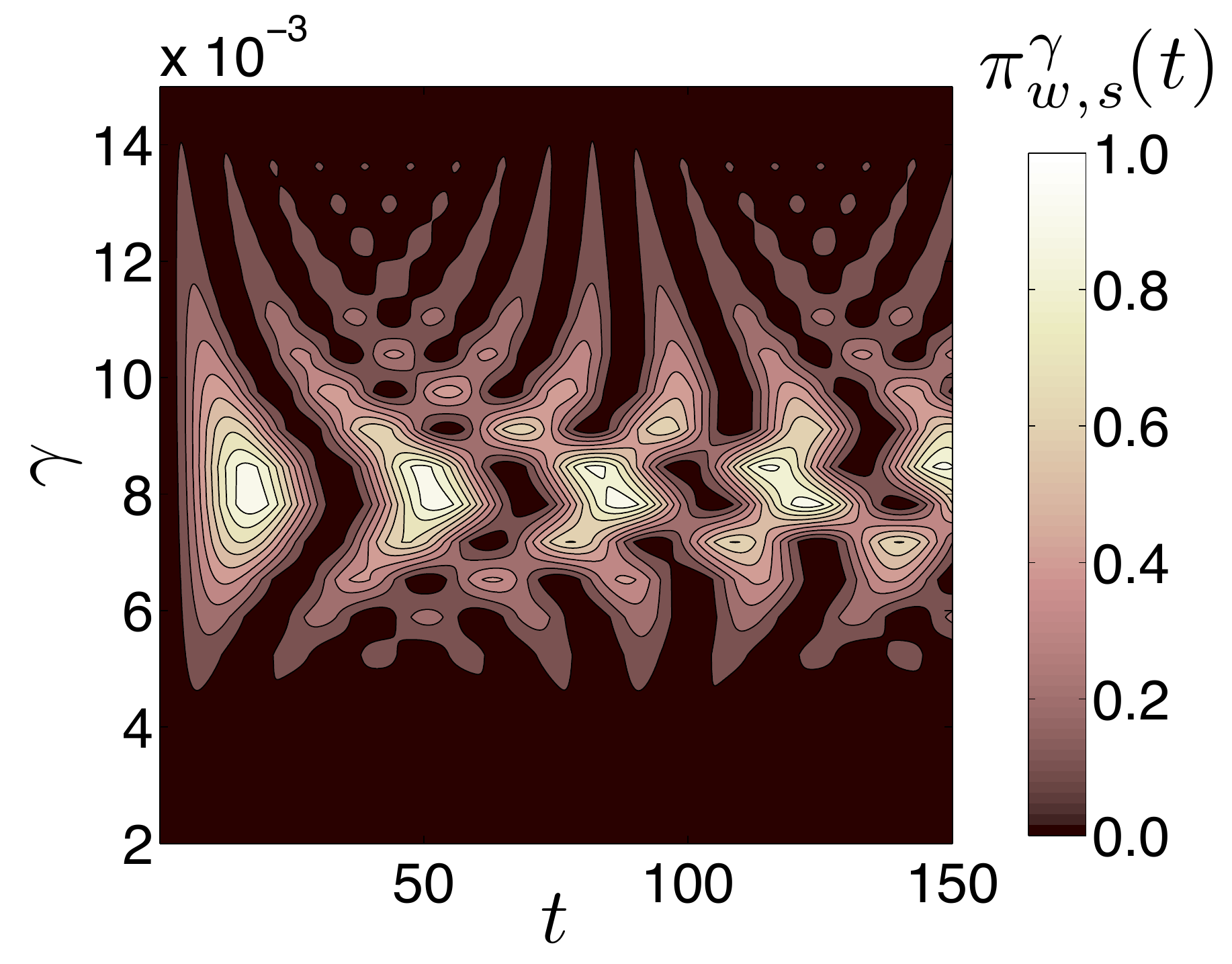}
\includegraphics[width=85mm,height=70mm]{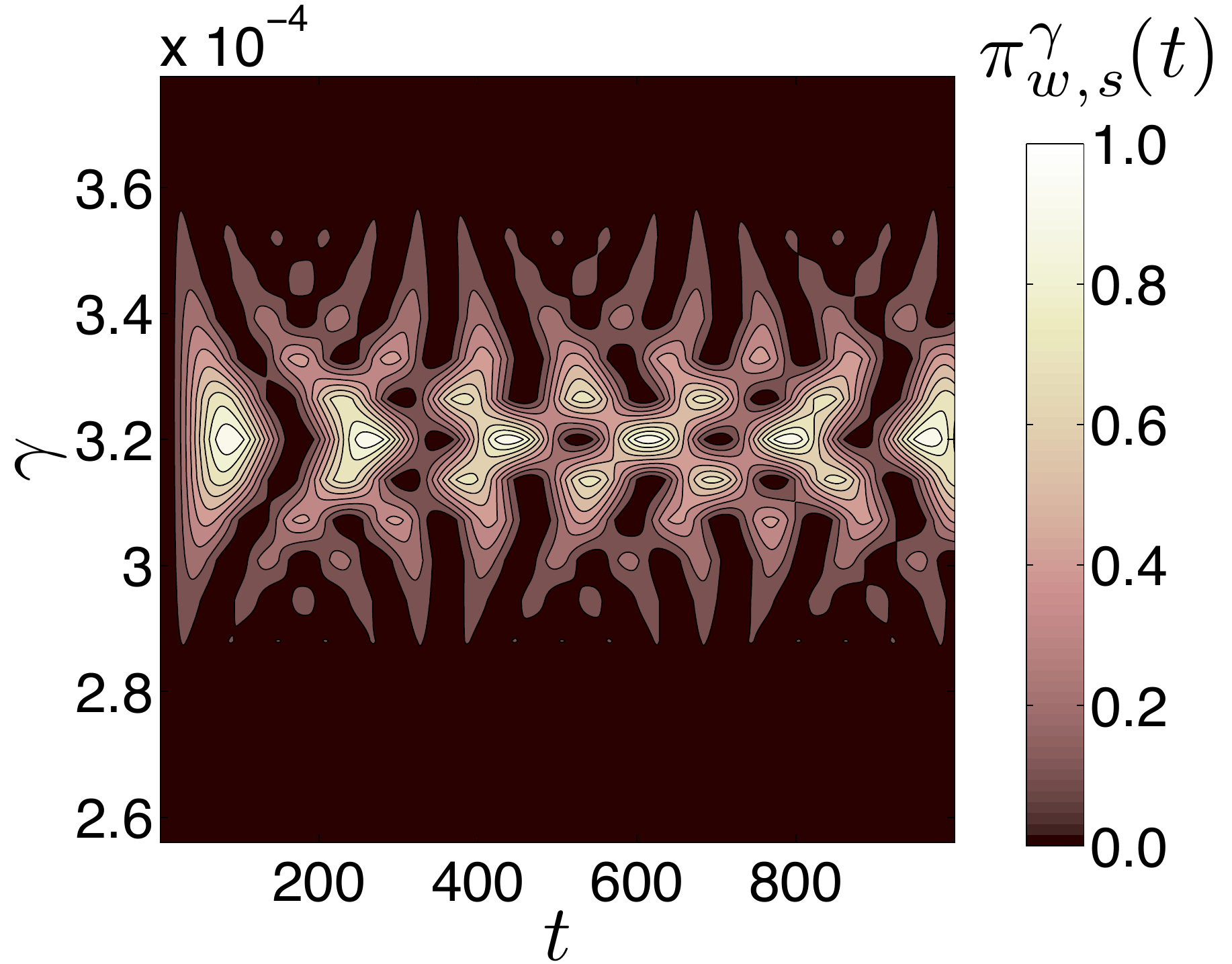} \caption{
\label{fig:Complete}(Color online) Contour plot of the success probability
$\pi_{w,s}(t,\gamma)$ as a function of (the dimensionless) time $t$ and of $\gamma$ for the complete
graph of size $N=124$ (top) and $N=3125$ (bottom). One can notice that for
$\gamma N =1$, namely $\gamma=8.1 \cdot 10^{-3}$ (top) and $\gamma=3.2 \cdot
10^{-4}$ (bottom), $\pi_{w,s}^{\gamma}(t)$ has a period $\tau=\pi
\sqrt{124}\approx 35$ and $\tau \approx 176$, respectively.} \end{figure}

We start our analysis from the complete graph $K_N$ for which, as shown in
\cite{childs}, at $\tilde{\gamma}=1/N$ the ground state changes sharply
from $|s\rangle$ to $|w\rangle$. This transition takes place at $t = \pi
\sqrt{N}/2$.  Due to the special topology of $K_N$, we are able to
calculate $\pi_{w,s}^{\gamma}(t)$ exactly, obtaining
\begin{equation}\label{eq:success_completo}
\pi_{w,s}^{\gamma}(t) = \frac{1}{N} \left[ 1 + \frac{4 (N-1) \sin^2{(t(\sqrt{4
\gamma + (N \gamma -1)^2}/2)}}{4+ \gamma( N -1/\gamma)^2 } \right], 
\end{equation}
see Appendix \ref{appe3} for details.

In Fig.~\ref{fig:Complete} we show $\pi_{w,s}(t,\gamma)$ for the complete
graph with $N=124$ and $N=3125$. We evaluated the figures both
numerically, by first diagonalizing $\mathbf{H}$ in
Eq.~(\ref{eq:success_prob}) and projecting on the states $| w \rangle$ and
$| s \rangle$ and also by making use of Eq.~(\ref{eq:success_completo}).
The results are numerically indistinguishable.  From
Fig.~\ref{fig:Complete} we see that, around the values $\gamma =  8 \times
10^{-3}$ and $\gamma = 3.2 \times 10^{-4}$ for $N=124$ and $N=3125$
respectively, $\pi_{w,s}^{\gamma}(t)$ reaches values very close to $1$. In
fact, analyzing Eq.~(\ref{eq:success_completo}) (see Appendix
\ref{appe3}), one finds that $\pi_{w,s}^{\gamma}(t)$ attains its maximal
value of $1$ for $\gamma N =1$ and for $t=\pi \sqrt{N} (k+1/2)$, where $k
\in \mathbb{Z}$. Therefore, $\gamma_{\mathrm{max}} = 1/N$. Furthermore,
due to the fact that the period between maxima is $T= \pi \sqrt{N}$ for
$\gamma_{\mathrm{max}}$, it follows that the CTQW takes $O(\sqrt{N})$
queries to find the target, in agreement with previous results
\cite{childs,farhi1}. On the other hand, the exact dependence on $t$ and
on $\gamma$ also allows to highlight the oscillating behaviour of
$\pi_{w,s}^{\gamma}(t)$. This means that, although we properly select
$\gamma_{\mathrm{max}}=1/N$, the result for the walk depends sensitively
also on $t$. In particular, for $t=k \pi \sqrt{N}, k \in \mathbb{Z}$, the
success probability is minimal and equals $1/N$ (which also corresponds to
the absolute minimum).

For the hypercubic lattices the overlaps of the states $|s\rangle$ and
$|w\rangle$ with $|\psi_0 \rangle$ and $|\psi_1 \rangle$ have already been
analyzed in \cite{childs}; here we display analogous plots (but for larger
sizes) in order to compare them with the corresponding success probability
$\pi_{w,s}^{\gamma}(t)$.
\begin{figure} \includegraphics[width=80mm,height=70mm]{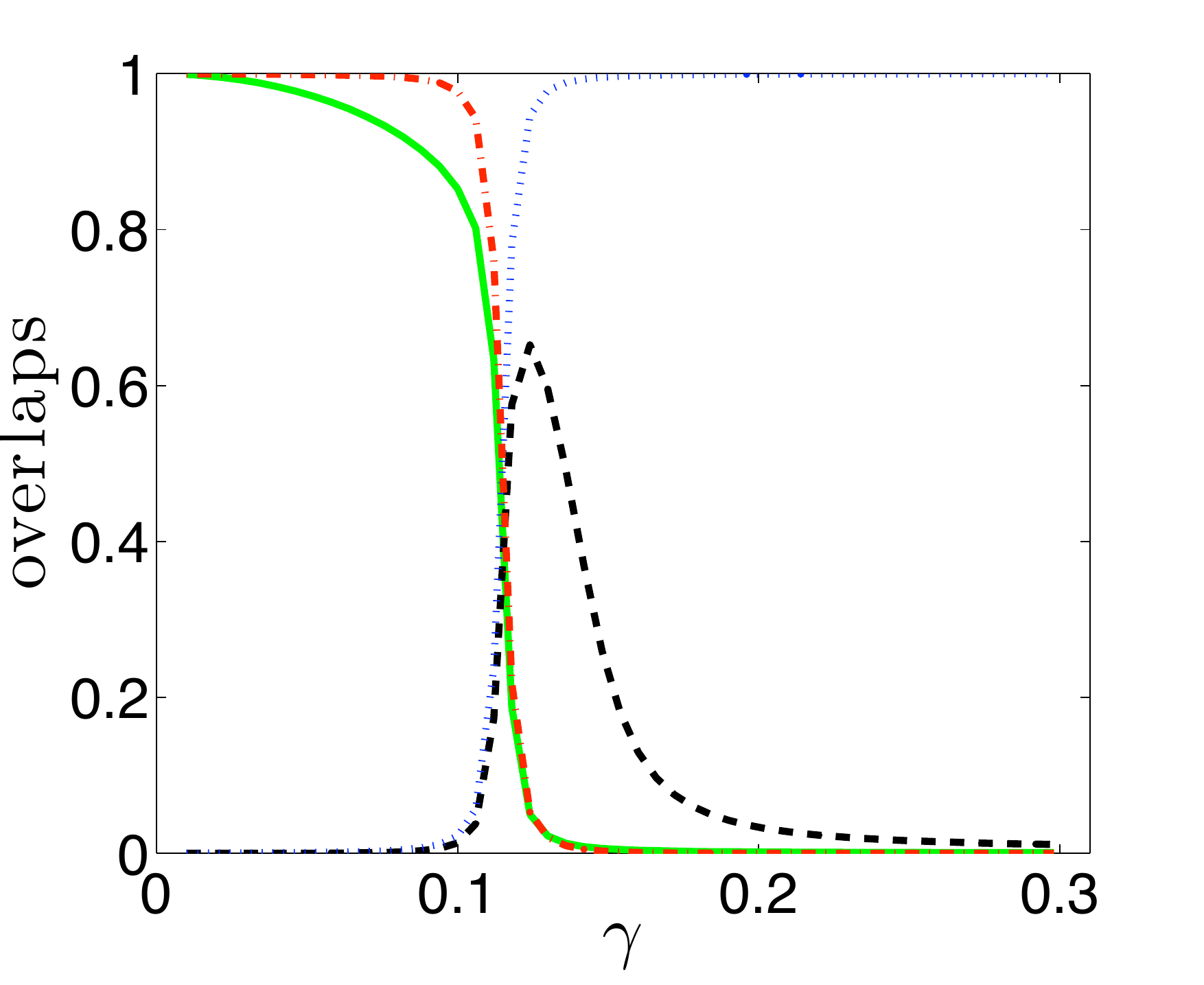}
\includegraphics[width=85mm,height=70mm]{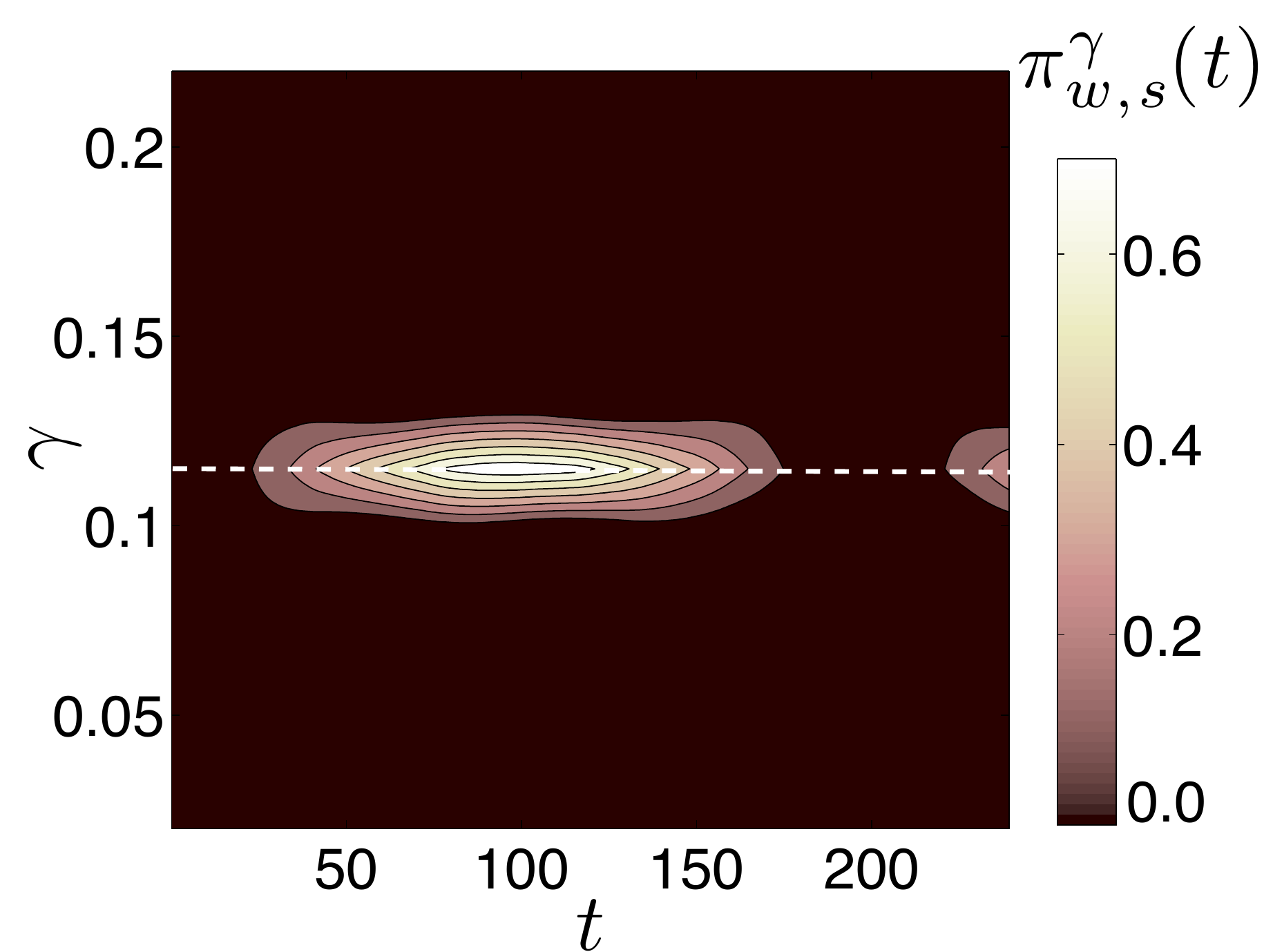} \caption{
\label{fig:Torus5}(Color online) $5$-dimensional torus with linear size $L=5$;
Top: Overlaps (symbols are as in Fig.~\ref{fig:DS_ext}); Bottom: Contour plot of
the success probability $\pi_{w,s}(t,\gamma)$ as a function of $t$ and of
$\gamma$; the dashed white line represents $\tilde{\gamma} \approx 0.12$.}
\end{figure}
In Fig.~\ref{fig:Torus5} we consider the case of a $5$-dimensional torus (i.e. a five-dimensional cubic lattice with periodic boundary conditions)
of linear size $L \equiv N^{1/d}=5$: The transition at $\tilde{\gamma}
\approx 0.12$ is very clear (see the top panel) and the success
probability is sharply peaked just at $\gamma_\mathrm{max} \approx
\tilde{\gamma}$ (see the bottom panel).

Away of the critical point $\tilde{\gamma}$ the success probability
quickly decays as a function of $\gamma$: it is just in the region of
largest overlap between the initial state $| s \rangle$ and the target
state $|w\rangle$ (namely around $\tilde{\gamma}$) that one expects an
optimal success probability.

Again, we notice that $\pi_{w,s}^{\gamma}(t)$ oscillates in time; for
$L=5$ and $\gamma = \gamma_{\mathrm{max}}$ the success probability ranges
from about $0$ to about $0.8$.  Moreover, for a given time $t$,
$\pi_{w,s}^{\gamma}(t)$ decays very fast as $|\gamma-\gamma_\mathrm{max}|$
increases. For instance, $\pi_{w,s}^{0.9
\gamma_\mathrm{max}}(70)/\pi_{w,s}^{\gamma_\mathrm{max}}(70)\approx0.05$.
As a result, the computational procedure for this structure can be very
efficient, provided that the parameter $\gamma$ can be sensitively
controlled.

\begin{figure} \includegraphics[width=80mm,height=70mm]{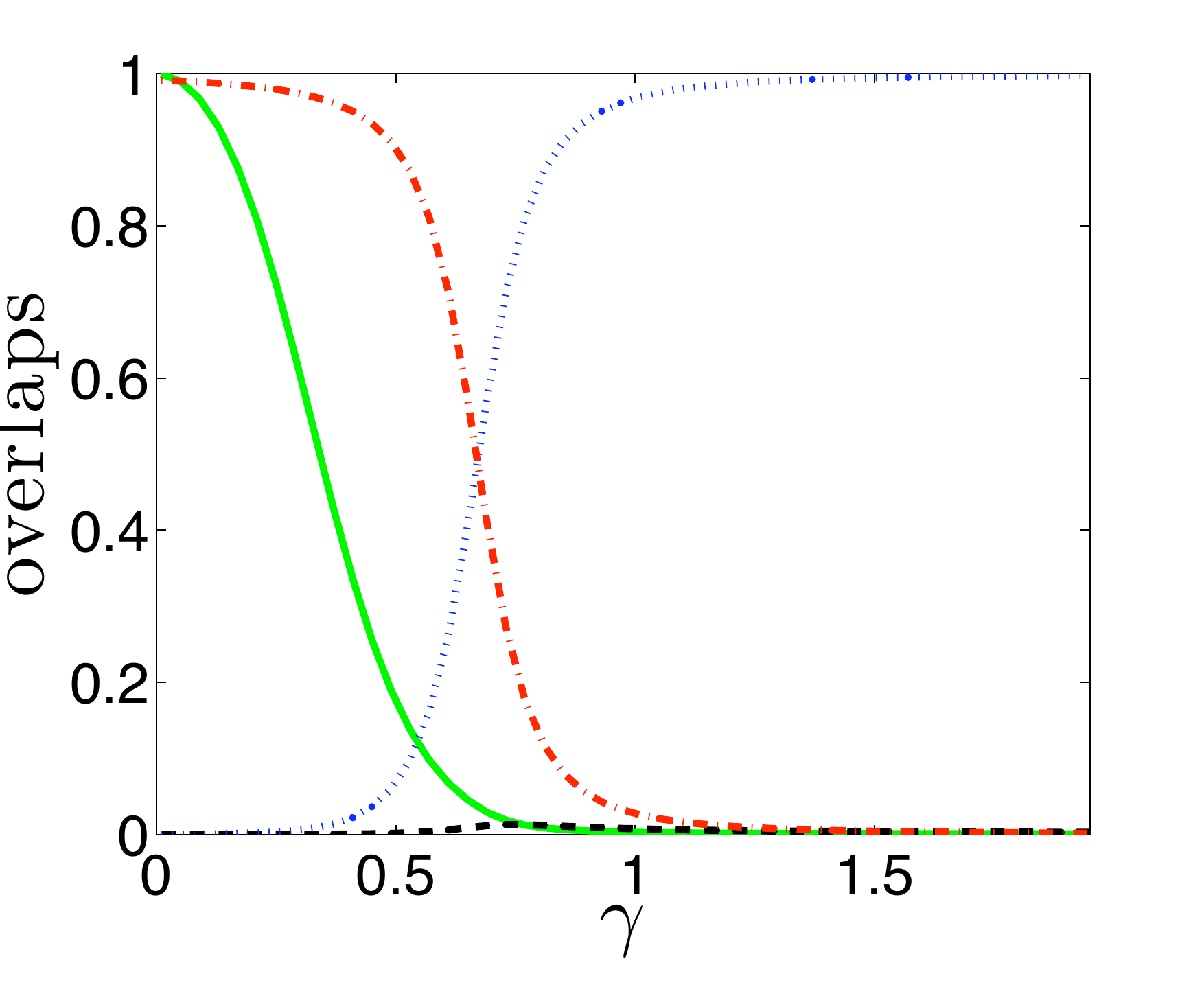}
\includegraphics[width=85mm,height=70mm]{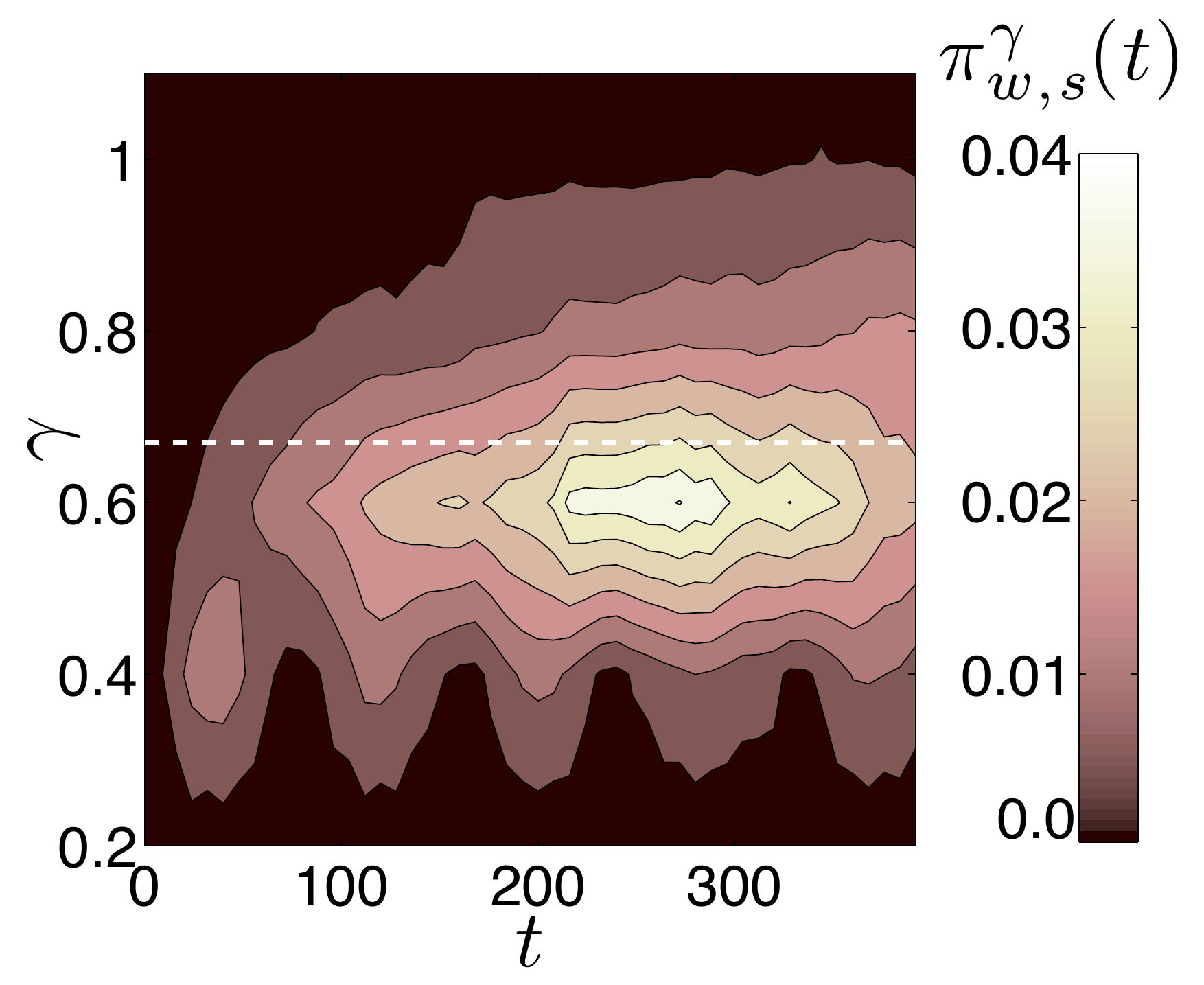}
\caption{\label{fig:Torus2}(Color online) $2$-dimensional square torus of linear
size $L=56$. Top: Overlaps (symbols are as in Fig.~\ref{fig:DS_ext}); Bottom:
Success probability as a function of time and $\gamma$; the dashed white line
represents $\tilde{\gamma}\approx 0.67$.} \end{figure}

For hypercubic lattices of dimension $d=4$, $d=3$ and $d=2$ (only the
latter case is depicted in Fig.~\ref{fig:Torus2}) the peaks get more and
more broadened and are of smaller magnitude. Notice that the low peaks
obtained are in agreement with the analytical results found in
\cite{childs} which predict for the $2$-dimensional torus a vanishing
success probability for $N \rightarrow \infty$. 

\begin{figure} \includegraphics[width=85mm,height=70mm]{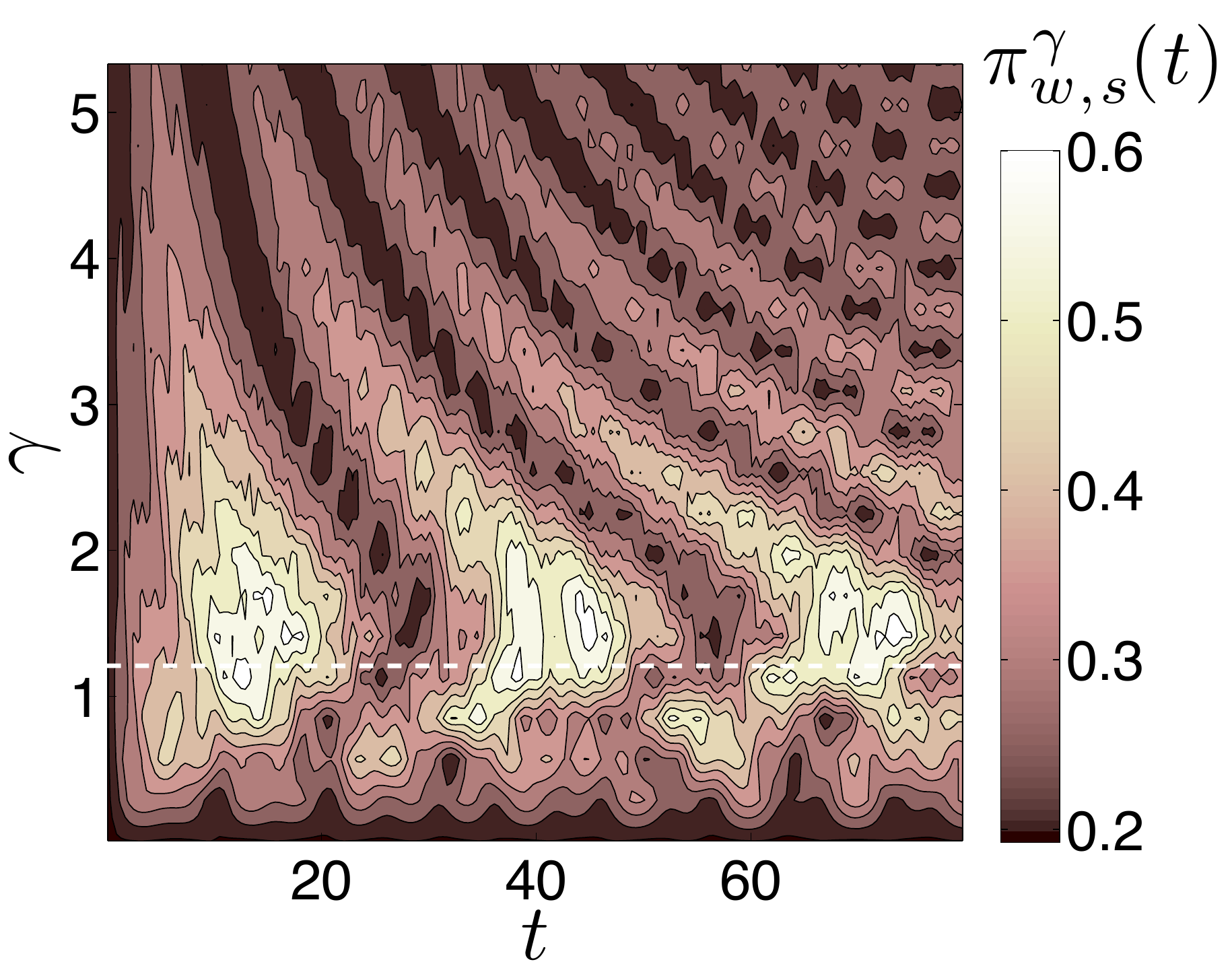}
\includegraphics[width=85mm,height=70mm]{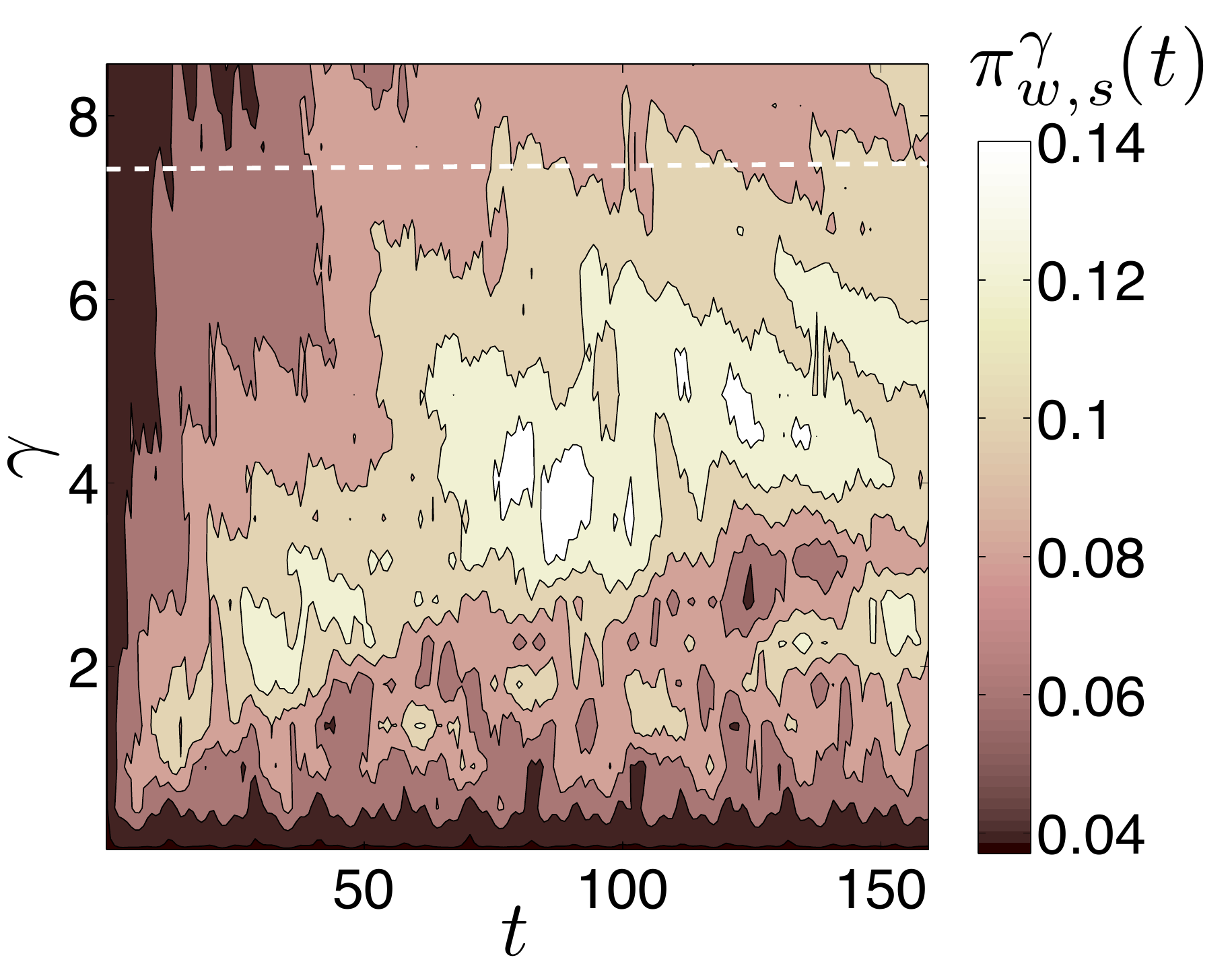} \caption{
\label{fig:ampl_DS}(Color online) Contour plot of $\pi_{w,s}^{\gamma}(t)$ for
the DSG of generation $g=3$ (top) and $g=6$ (bottom); the ``critical'' values
$\tilde{\gamma}$ are represented by the dashed line: $\tilde{\gamma}(3)\approx
1.20$ and $\tilde{\gamma}(6)\approx 7.38$ (see Fig.~\ref{fig:DS_ext}).}
\end{figure}

\begin{figure} \includegraphics[width=85mm,height=70mm]{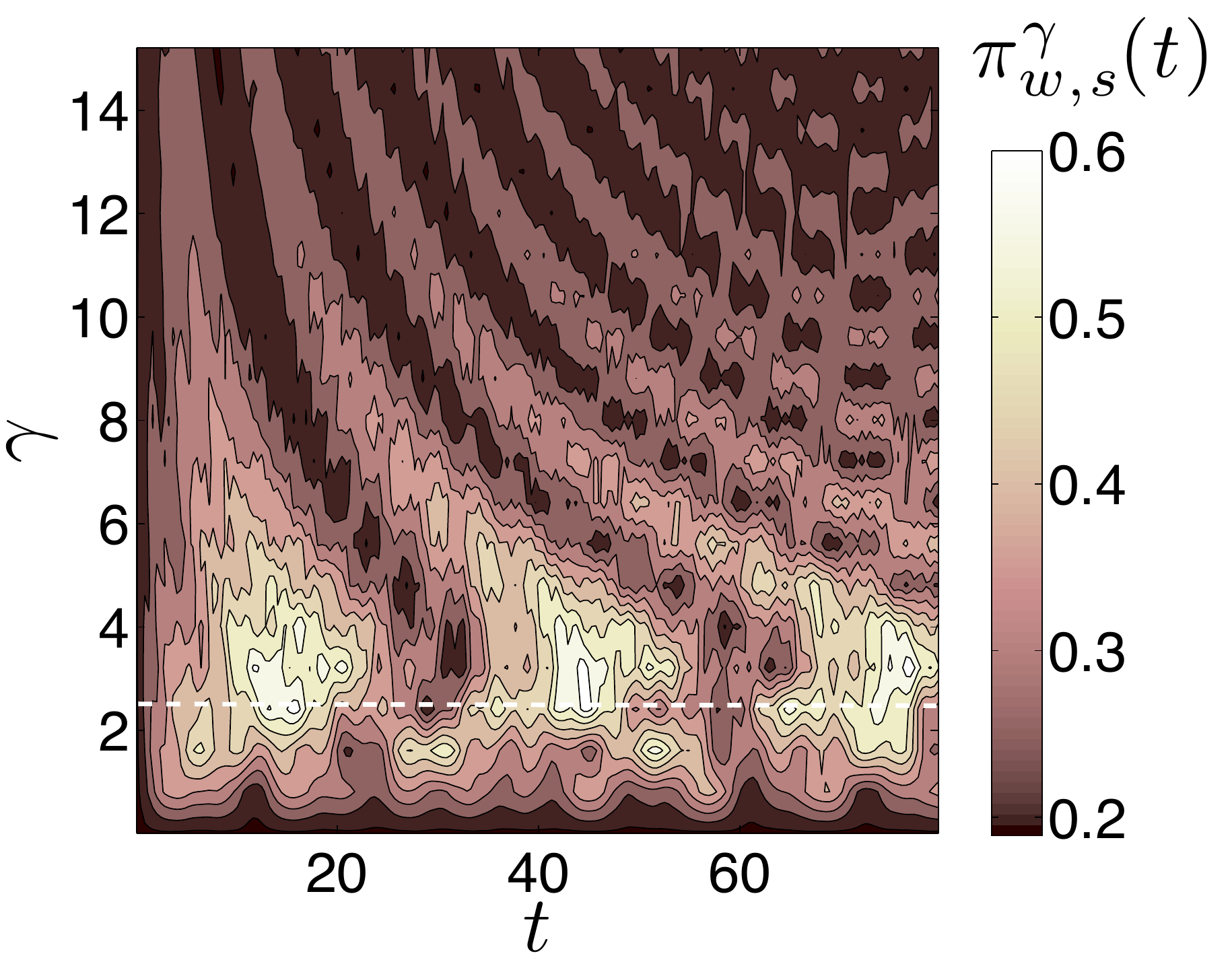}
\includegraphics[width=85mm,height=70mm]{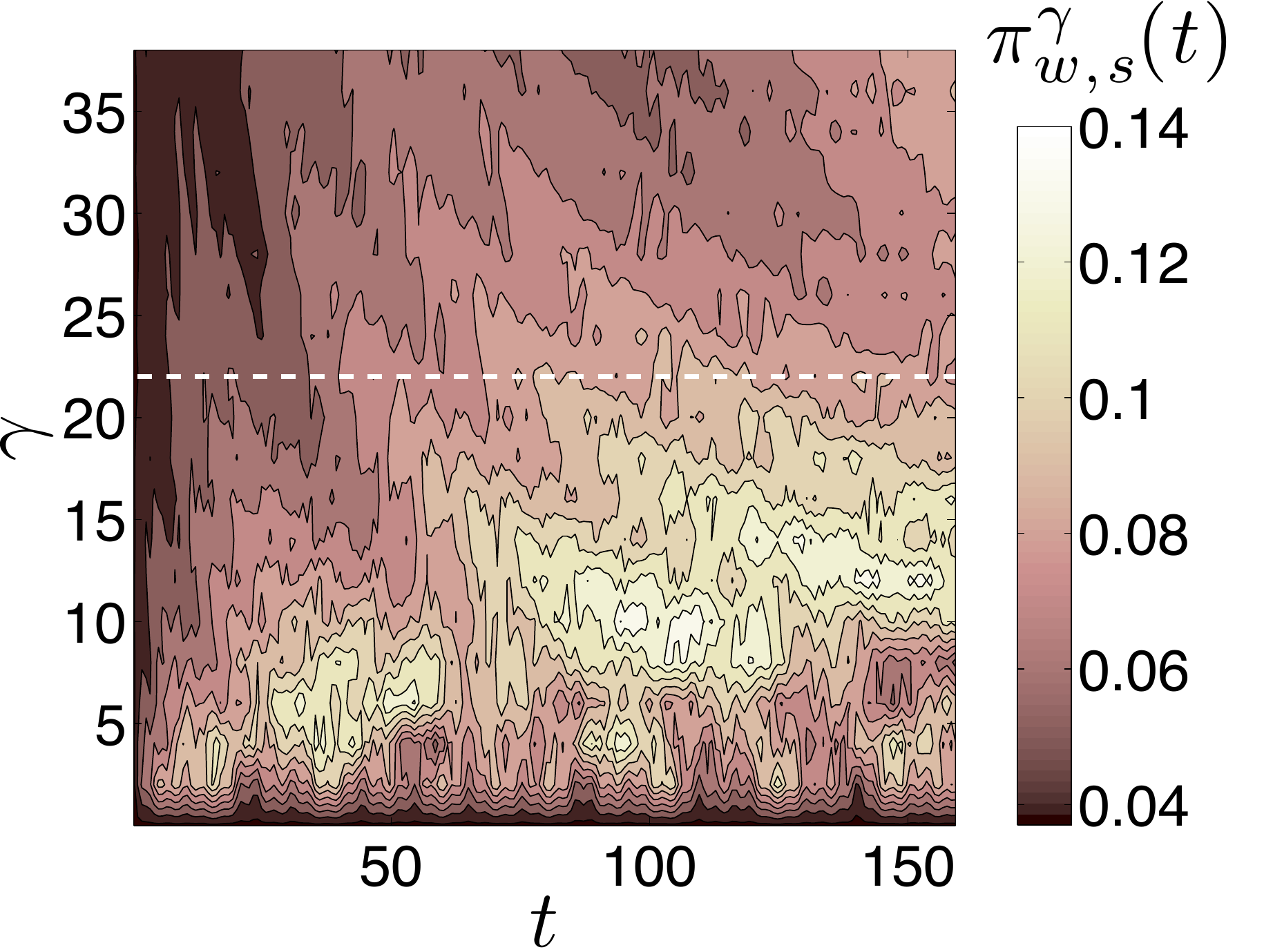}
\caption{\label{fig:ampl_T}(Color online) Contour plot of
$\pi_{w,s}^{\gamma}(t)$ for the TF of generation $g=3$ (top) and $g=6$ (bottom);
notice that $\tilde{\gamma}(3)\approx2.59$ and $\tilde{\gamma}(6)\approx21.31$
(see Fig.~\ref{fig:T_ext}).} \end{figure}

We now turn to the analysis of the success probability for the DSG, the TF and
the CT, represented in Figures \ref{fig:ampl_DS}, \ref{fig:ampl_T} and
$12$, respectively; the upper panels display the situation for
small, the lower panels for larger networks.  First of all, we notice that, as
previously found for regular lattices \cite{childs}, also for DGS, TF and CT,
$\pi_{w,s}^{\gamma}(t)$ exhibits peaks which are lower and lower as the size $N$
is enlarged. Moreover, for small sizes (upper panels) $\gamma_{\mathrm{max}}
\approx \tilde{\gamma}$, namely the maxima for the success probability occur for
values of $\gamma$ which are approximately equal to $\tilde{\gamma}$. On the
other hand, for large sizes (lower panels), in the temporal range considered
here, $\tilde{\gamma}$ provides an upper bound for $\gamma_{\mathrm{max}}$.
However, the most striking feature which emerges from the comparison between the
contour plot of the success probability for translationally invariant structures (see
Figs.~\ref{fig:Torus5} and \ref{fig:Torus2}) and for fractal/hierarchical structures (see
Figs.~ \ref{fig:ampl_DS}-$12$) is that for the latter
$\pi_{w,s}(t)$ is much more rough and broadened.

\begin{figure} \label{fig:ampl_CT}
\includegraphics[width=85mm,height=70mm]{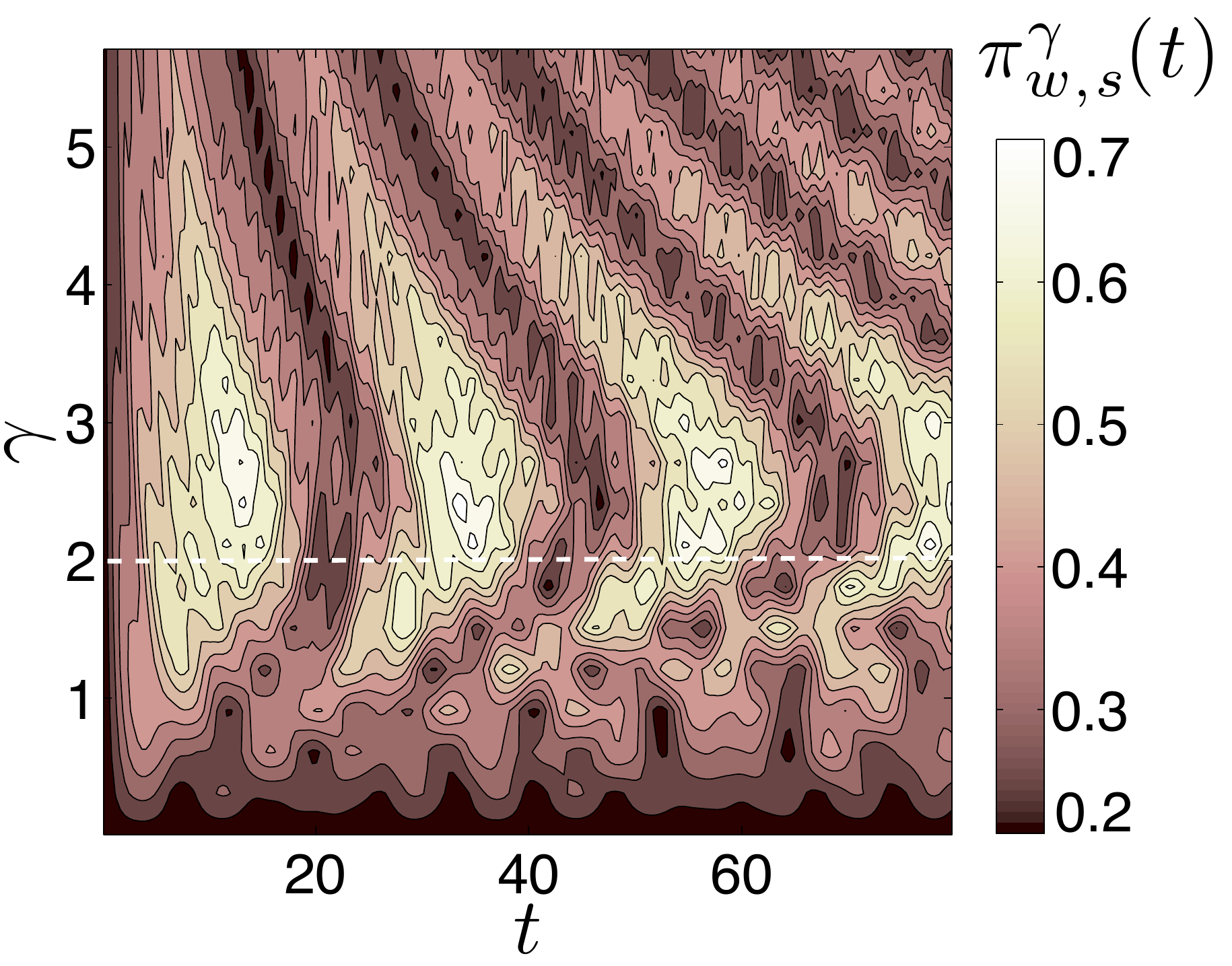}
\includegraphics[width=85mm,height=70mm]{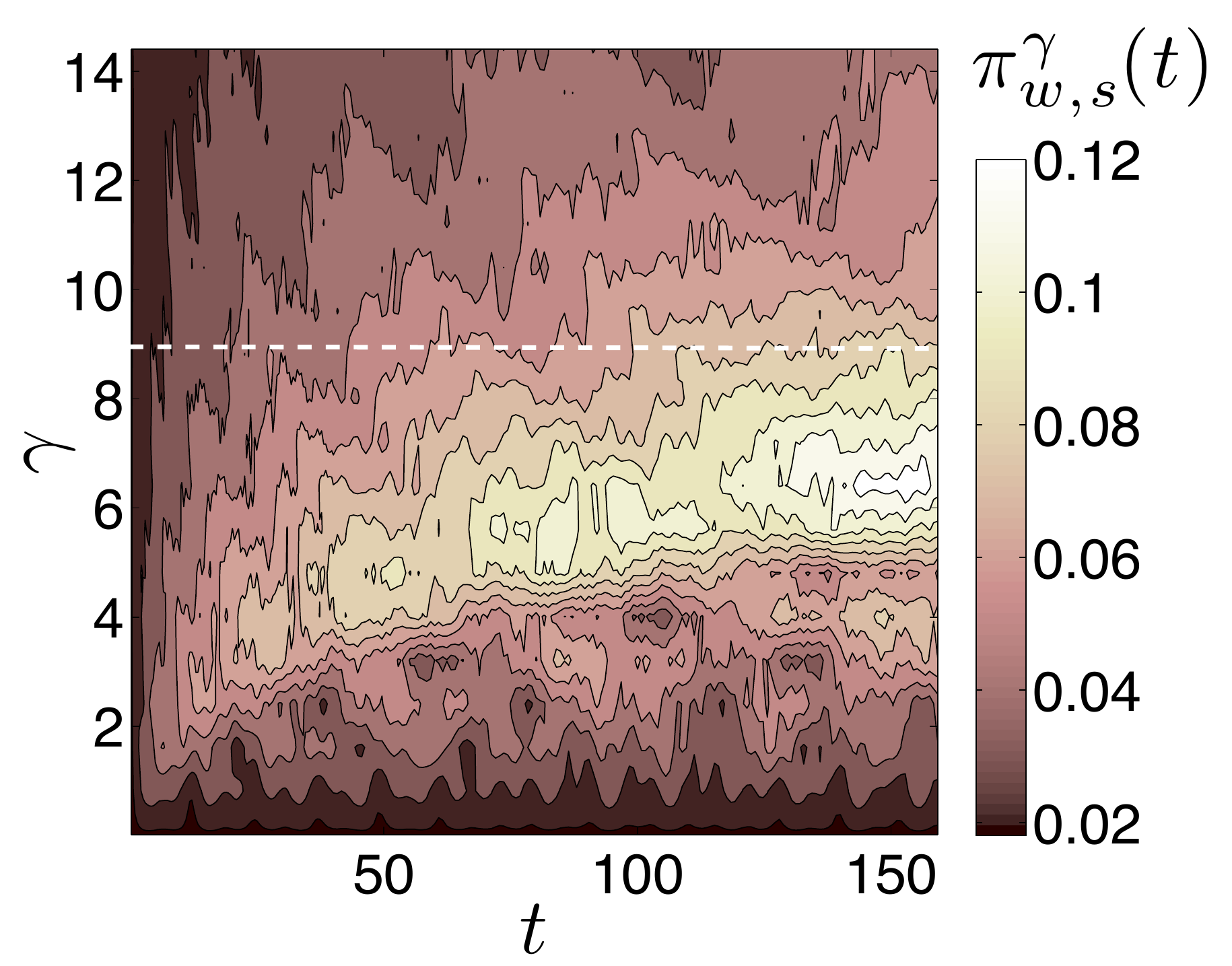}
\caption{(Color online) Contour plot of
$\pi_{w,s}^{\gamma}(t)$ for the CT of generation $g=3$ (top) and $g=10$
(bottom); notice that $\tilde{\gamma}(3)\approx 2.06$ and
$\tilde{\gamma}(10)\approx 8.87$ (see Fig.~\ref{fig:CT_ext}).} \end{figure}

Now, it is worth comparing the $5$-dimensional torus of
Fig.~\ref{fig:Torus5}, the square torus of Fig.~\ref{fig:Torus2} and the
CT of Fig.~$12$ since all three are of comparable size $N$.
From the computational point of view the $5$-dimensional torus corresponds
to the best situation: $\pi_{w,s}^{\gamma}(t)$ is sharp and reaches its
maximum value around $0.8$ after approximately $100$ unit steps; the CT
displays a success probability around $0.12$ after $150$ time steps; the
$2$-dimensional torus corresponds to an ineffective candidate situation:
after $300$ time steps the peak is still less than $0.1$. 

In order to compare also to structures with spectral dimensions larger than
four, we show in Fig.~$13$ the success probabilities
$\pi_{w,s}^{\gamma}(t)$ for the cartesian products of a DSG of generation $g=3$
and a square lattice of size $L=8$ as well as a cubic lattice of size
$L=8$. Although the maxima of the success probabilities are in both cases
larger than the ones for the structures with low dimensions, they show
still a fairly unregular pattern. This is in contrast to the highly
regular structure of the 5-dimensional torus, see Fig.~\ref{fig:Torus5}.

\begin{figure} \label{fig:ampl_CP_DSGL}
\includegraphics[width=85mm,height=70mm]{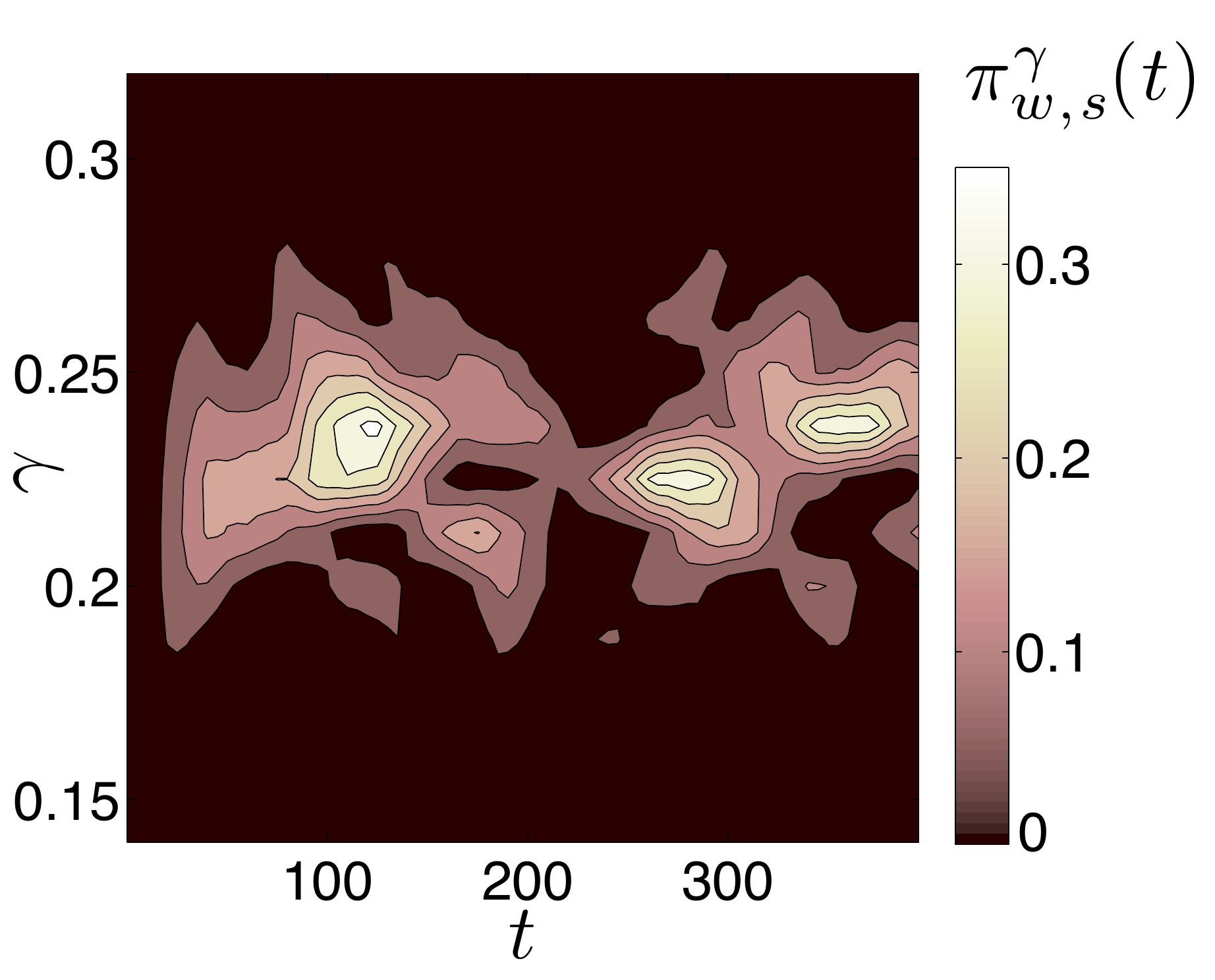}
\includegraphics[width=85mm,height=70mm]{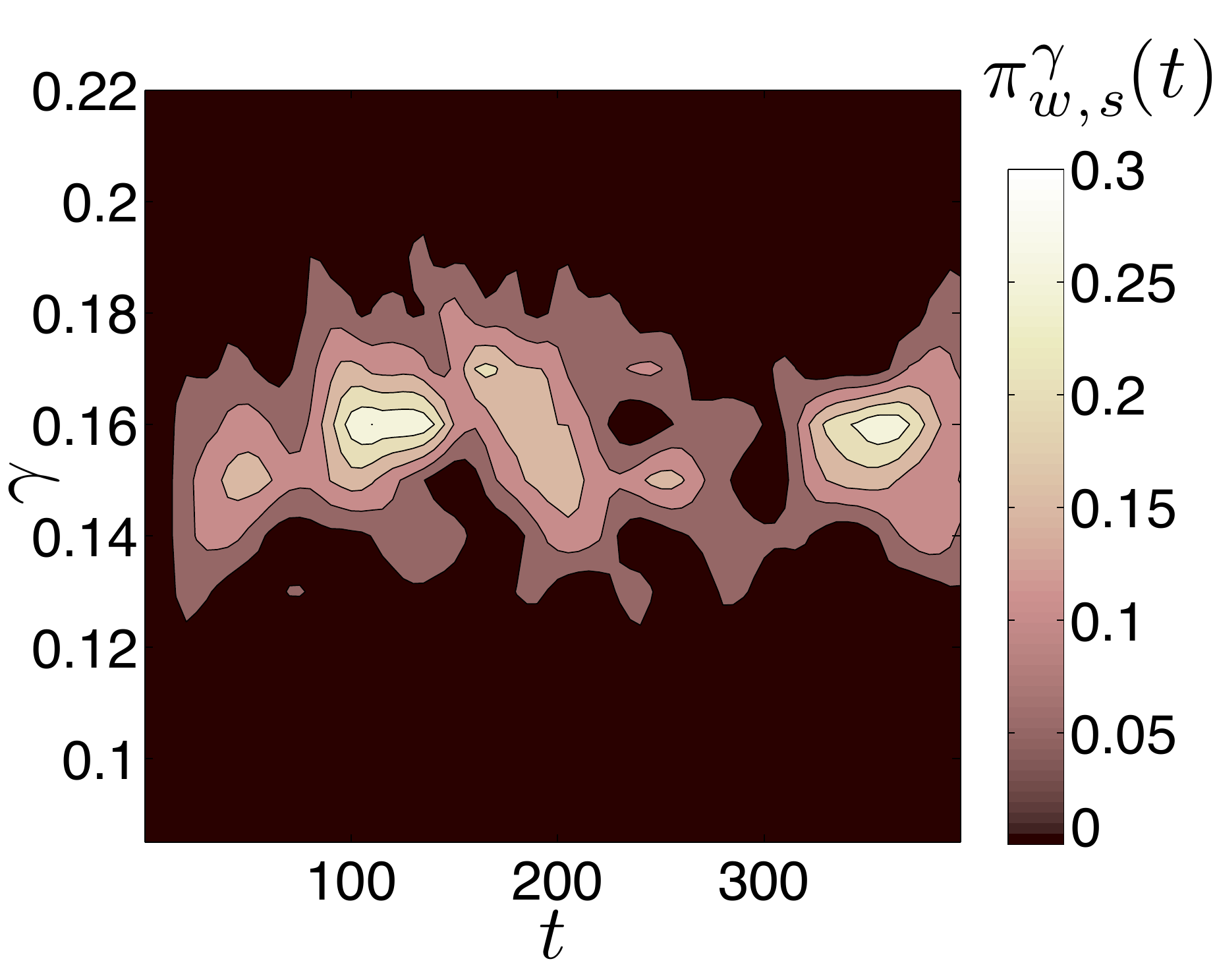}
\caption{(Color online) Contour plot of
$\pi_{w,s}^{\gamma}(t)$ for the cartesian product of a DSG of generation $g=3$
and a square lattice of size $L=8$ (top) and a cubic lattice of size $L=8$
(bottom).} \end{figure}

Finally, we stress that, as a result of interference phenomena,
$\pi_{w,s}^{\gamma}(t)$ oscillates with time. This has some important
consequences: Although we can determine and set the optimal
$\gamma_{\mathrm{max}}$, the probability of the CTQW reaching the target
depends on the instant of time at which it is calculated. In particular,
oscillations are ``faster'' for systems of smaller size; for example for
the complete graph we find a period $\tau(\gamma) = 2 \pi / \sqrt{(\gamma
N -1)^2 + 4 \gamma}$ (see Eq.~(\ref{eq:success_completo})), namely
$\tau(\gamma_{\mathrm{max}})= \pi \sqrt{N}$, while for the $5$-dimensional
torus the numerical analysis makes it possible to estimate a period $\tau$ which
grows exponentially with the lattice size $L$; for $L=5$ we get $\tau
\approx 200$ (see Fig.~\ref{fig:Torus5}).

\section{Conclusions and perspectives}\label{sec:conclusions} 

In this work we considered CTQWs mimicking Grover's
quantum search problem; we especially focused on how the topology of the
space over which the walk takes place affects the position and sharpness of the transition of the ground state.
Previous studies \cite{childs} highlighted that for translationally invariant
structures, such as the hypercubic lattices, the quantum walk can be highly
efficient for sufficiently high dimensions,
i.e. $d>4$. However, here we evidence that on generic graphs the dimension
does not represent the key geometric parameter; indeed, both the (average) coordination number and the fact
that the structure is translationally invariant or not determine the sharpness of the transition.
In fact, on the one hand, a high coordination reduces the distance among
nodes and increases the possibility of interference effects, on the other
hand, (in the absence of a target) translational invariance prevents the emergence of localization
effects \cite{ABM2008}.

In particular, we considered the success probability
$\pi^{\gamma}_{w,s}(t)$ (here $|s\rangle$ and $|w\rangle$ are the initial
and the target state, respectively) as a function of the computation time
$t$ and of a properly tunable parameter $\gamma$ entering the Hamiltonian.
We showed that for highly dimensional ($d>4$)
translationally invariant structures ($5$-dimensional torus) there exists a
narrow range of $\gamma$ around a specific value
$\tilde{\gamma}$ where the ground state $| \psi_0 \rangle$ undergoes a
transition from having a large overlap with $|s\rangle$ to a state with a
large overlap with $|w\rangle$. This corresponds to a sharply peaked
success probability: $\pi^{\gamma}_{w,s}(t)$ displays a set of maxima just
at $\tilde{\gamma}$; the first one is reached after $O(\sqrt{N})$ queries.
Conversely, for structures with low coordination number and/or fractal or
hierarchical topology - such as the cubic ($L_3$) and square ($L_2$) tori,
the DSG, the TF, the CT
and the Cartesian products DSG $\times$ $L_2$ and DSG $\times$ $L_3$ - the
transition from the initial state to the target state takes place over a
wider region of $\gamma$ around the value $\tilde{\gamma}$.  As a
consequence of such a spread-out transition, the success probability
displays broadened peaks whose locations depend on time. Therefore, for
the non-translationally invariant structures considered here, even with large
fractal dimension ($\tilde{d} > 4$), the large success probabilities found for
high-dimensional periodic lattices are not recovered.
These results, in agreement with previous findings \cite{childs}, highlight a possible connection between the sharpness of the transition occurring at $\tilde{\gamma}$ and the efficiency of the search algorithm. A mathematical, rigorous proof stating whether a sharp transition is a necessary condition for a good algorithm, which is beyond the aim of this article, could provide a very useful tool for further investigations on quantum search algorithms.

For the DSG we also proved that $\tilde{\gamma}$ scales like
$N^{2/\tilde{d}+\alpha}$ ($-1 \leq \alpha < 0$); interestingly our results
suggest that such a scaling might be generalized to all (exactly
decimable) fractals with spectral dimension $\tilde{d} < 2$.

Apart from the deterministic fractals considered here, it will be
extremely interesting to also consider disordered structures such as
percolation clusters and random graphs characterized by a degree
distribution $P(z)$. These networks display a tunable average degree
$\langle z \rangle$, which, in principle, can take values ranging from $0$
for totally disconnected networks up to $N$ for completely connected
networks. According to the results discussed here, we expect that for a
sufficiently large $\langle z \rangle$ and for a sufficiently peaked
$P(z)$ the transition from the ground state $|s \rangle$ to a state with
significant overlap with $|w \rangle$ occurs sharply and that,
consequently, the CTQW speeds up.

\section*{Acknowledgements} Support from the Deutsche Forschungsgemeinschaft
(DFG) and the Fonds der Chemischen Industrie is gratefully acknowledged. EA
thanks the Italian Foundation ``Angelo della Riccia'' for financial support.

\appendix \section{Generic structures}\label{appe1} Let us denote by $\{ |
\phi_k \rangle \}$ and by $\{ \mathcal{E}(k)\}$ the sets of eigenstates and of
eigenvalues of the Laplacian $\mathbf{L}$, respectively. On the basis of the
eigenstates the state $|w \rangle$ localized at the target site can be written
\begin{equation} 
|w \rangle =  \sum_{k=0}^N a_k |\phi_k
\rangle.  
\end{equation} 
Now, let us consider the complete Hamiltonian
$\mathbf{H}$ and the corresponding eigenvalue equation for the state labeled
$a$: 
\begin{equation} \label{eq:eigenvalue} 
\mathbf{H} |\psi_a \rangle = (\gamma
\mathbf{L} - |w \rangle \langle w |) |\psi_a \rangle = E_a |\psi_a \rangle.
\end{equation} 
As shown in \cite{childs}, when one sets $R_a = |\langle w|\psi_a
\rangle |^2$, it is possible to write 
\begin{equation} \label{eq:eigenvalue1}
|\psi_a \rangle =  \frac{\sqrt{R_a}}{\gamma \mathbf{L} - E_a }|w \rangle,
\end{equation} 
and 
\begin{equation} \label{eq:normalize} 
F(E_a) \equiv \left
\langle w \left| \frac{1}{\gamma \mathbf{L} - E_a } \right| w \right \rangle   =
1.  
\end{equation} 
Here we notice that Eq.~\ref{eq:eigenvalue1}, and therefore
also Eq.~\ref{eq:normalize}, hold when the eigenvalue $E_a$ is
\textit{non-degenerate} (for example, this condition is not fulfilled if we
place the trap on the central node of the TF and CT). Equations
(\ref{eq:eigenvalue1}) and (\ref{eq:normalize}) allow then to express the
overlap of a given state $|\psi_a \rangle$ (with non-degenerate $E_a$) with $|s
\rangle$ as 
\begin{equation} \label{eq:overlap} 
| \langle s |  \psi_a \rangle |^
2 =  \frac{1}{E_a^2 F'(E_a) N} = \frac{R_a}{N E_a^2}, 
\end{equation} 
where
\begin{equation} \label{eq:F_deriv} 
F'(E) \equiv \frac{\partial F(E)}{\partial
E}=  \left \langle w \left| \frac{1}{(\gamma \mathbf{L} - E)^2} \right| w \right
\rangle, 
\end{equation} 
and, in particular, 
\begin{equation} 
F'(E_a) =  \left \langle w \left| \frac{1}{(\gamma \mathbf{L} - E_a)^2} \right|
w \right \rangle = \frac{1}{R_a}, 
\end{equation} 
(see \cite{childs} for more
details).  It is worth underlining that Eq.~(\ref{eq:overlap}) found in
\cite{childs} for hypercubic lattices actually holds for all structures. The
topological details are contained in $E_a$ and $F'(E_a)$.

We now proceed with the calculations without making any assumptions on the
topology of the database. First, by using Eq.~(\ref{eq:expansion}) and $a_k
a_k^{*} = |a_k|^2$, we rewrite $F(E)$ as 
\begin{equation}
\label{eq:normalize_gen} 
F(E)= \sum_k \frac{a_k a_k^{*}}{\gamma \mathcal{E}(k) -
E}= \sum_k \frac{|a_k|^2}{\gamma \mathcal{E}(k) - E}, 
\end{equation} 
whose
derivative is 
\begin{equation} \label{eq:F_deriv_gen} F'(E) =  \sum_{k=1}^{N}
\frac{|a_k|^2}{(\gamma \mathcal{E}(k) - E)^2}.  
\end{equation} 
For a generic
index $k$, $0 \leq |a_k|^{2}  < 1$; while the lower bound is clear, the upper
bound derives from the fact that the eigenstate $|w\rangle$ can not correspond
to any Laplacian eigenstate.  For Euclidean lattices Bloch's theorem makes it possible to
write $\langle x | \phi_k \rangle = e^{i k x}/ \sqrt{N}$ so that
$|a_k|^{2}=1/N$, for any $k$.  On the other hand, for a generic connected
structure, a priori, one can only write $|a_0|^2 = 1/N$, as a consequence of the
fact that the Laplacian eigenstate corresponding to the smallest eigenvalue
$\mathcal{E}(0)=0$  is just $| \phi_0 \rangle \equiv | s \rangle$. This suggests
a proper restriction of the previous upper bound: $\max_{k \neq 0} |a_k|^2 \leq
C N^{\alpha}$, with $C >0$ and $-1 \leq \alpha < 0$, both depending on the
particular topology chosen.  In particular, for Euclidean structures, $C=1$ and
$\alpha=-1$ and one expects that the more inhomogeneous the topology, the larger
$\alpha$.  By means of numerical calculations we can estimate $\alpha$: for not
too small DSG and TF we find that $\alpha \approx -0.9$ (see
Fig.~$14$).

\begin{figure} \label{fig:alfa} \includegraphics[height=60mm]{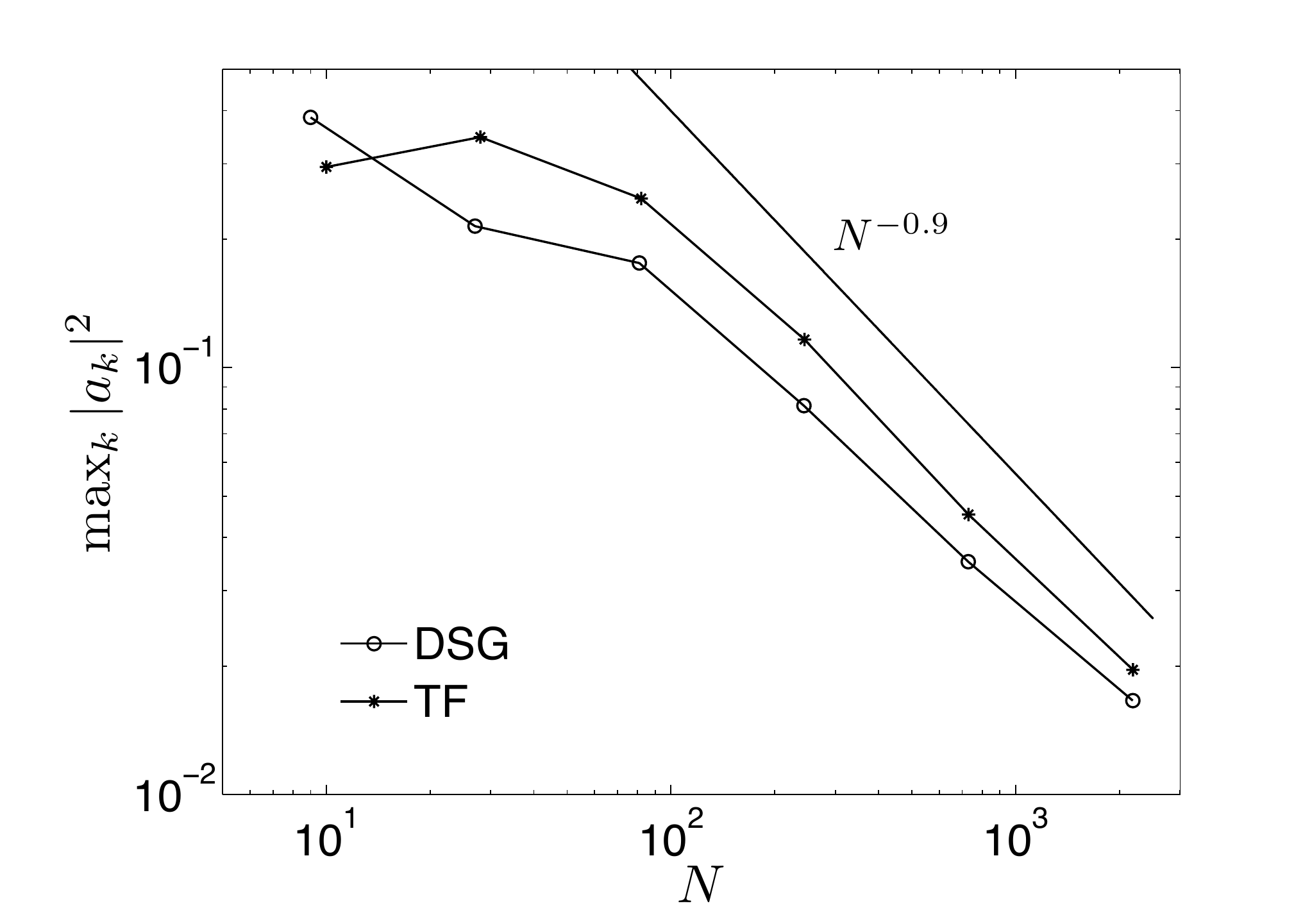}
\caption{Numerical estimate of $\alpha$ for the DSG and the
TF.} \end{figure}

Now, before going on, we define the quantity $\xi_j$, which will be useful in
the following: 
\begin{equation} \label{eq:xi_j} 
\xi_j \equiv \sum_{k \neq 0}
\frac{|a_k|^{2}}{[\mathcal{E}(k)]^j} \leq C N^{\alpha}  \sum_{k \neq 0}
\frac{1}{[\mathcal{E}(k)]^j}  \equiv C N^{\alpha} \zeta_j, 
\end{equation} 
where
we set $\sum_{k \neq 0} 1/[\mathcal{E}(k)]^j  \equiv \zeta_j$ and used the upper
bound $|a_k|^{2} \leq C N^{\alpha} < 1$; for hypercubic lattices $\xi_j =
\zeta_j /N$ which can be approximated by an integral \cite{childs}.

Using Eqs.~(\ref{eq:overlap}) and (\ref{eq:F_deriv_gen}) the overlap of $| s
\rangle$ with the ground state turns out to be 
\begin{eqnarray}
\label{eq:s_0_preli} \nonumber |\langle s | \psi_0 \rangle |^2 &=& \left [1 + N
E_0^2 \sum_{k \neq 0} \frac{|a_k|^{2}}{(\gamma \mathcal{E}(k) + |E_0|)^2} \right
]^{-1} \\ \nonumber &>& 1 - N E_0^2 \sum_{k \neq 0} \frac{|a_k|^{2}}{(\gamma
\mathcal{E}(k) + |E_0|)^2} \\ &>& 1 - \frac{N E_0^2}{\gamma^2} \sum_{k \neq 0}
\frac{|a_k|^2}{[\mathcal{E}(k)]^2}, 
\end{eqnarray} 
where in the first inequality
we used that the sum is positive, while in second inequality we used that both
$\mathcal{E}(k)$ and $|E_0|$ are positive. From Eq.~(\ref{eq:s_0_preli}) and
Eq.~(\ref{eq:xi_j}), we have 
\begin{equation} \label{eq:s_0} 
|\langle s | \psi_0
\rangle |^2  > 1 - \frac{N E_0^2}{\gamma^2} \xi_2.  
\end{equation}

In general, as $\gamma$ is varied, $E_0$ is bounded as $0 \leq |E_0| \leq 1$ and
the bounds can be improved by exploiting the following 
\begin{eqnarray}
\label{eq:E_0} \nonumber 1= F(E_0) = \frac{|a_0|^{2}}{|E_0|} + \sum_{k \neq 0}
\frac{|a_k|^{2}}{\gamma \mathcal{E}(k)+ |E_0|} \\ < \frac{1}{N |E_0|} + \sum_{k
\neq 0} \frac{|a_k|^{2}}{\gamma \mathcal{E}(k)}=\frac{1}{N
|E_0|}+\frac{\xi_1}{\gamma}.  
\end{eqnarray} 
In fact, we get 
\begin{equation}
\label{eq:bound_E0} \frac{1}{N} < |E_0| < \frac{1}{N} \frac{\gamma}{\gamma -
\xi_1}, 
\end{equation} 
where for the lower bound we used that the first sum
appearing in Eq.~(\ref{eq:E_0}) is positive and that $a_0 = a_0^{*}=1/\sqrt{N}$.

Now, from Eq.~(\ref{eq:s_0}) and Eq.~(\ref{eq:bound_E0}) it follows
straightforwardly that 
\begin{equation} \label{eq:bound_ground} 
1 - |\langle s |
\psi_0 \rangle |^2 <  \frac{\xi_2}{N (\gamma - \xi_1 )^2} \leq \frac{C
N^{\alpha-1} \zeta_2}{(\gamma - C N^{\alpha} \zeta_1)^2}.  
\end{equation}
Following analogous arguments we find for the first non-degenerate excited state
labelled as $1$, being $E_1$ non degenerate,
\begin{equation} 
|\langle s | \psi_1 \rangle |^2 > 1 - N E_1^2 \sum_{k \neq}
\frac{|a_k|^{2}}{(\gamma \mathcal{E}(k) - E_1)^2}, 
\end{equation} 
and
\begin{equation} 
E_1 < \frac{1}{N} \frac{\gamma}{(\xi_1 - \gamma)},
\end{equation} 
from which we get 
\begin{equation} \label{eq:bound_any} 1 -
|\langle s | \psi_1 \rangle |^2 <  \frac{\xi_2}{N (\xi_1 - \gamma )^2} \leq
\frac{C N^{\alpha-1} \zeta_2}{(C N^{\alpha} \zeta_1 - \gamma)^2}.
\end{equation} 
Notice that in this case $\gamma < \xi_1$ due to the fact that
$E_1 >0$.

Now, by comparing Eq.~(\ref{eq:bound_ground}) and Eq.~(\ref{eq:bound_any}) we
can evidence the existence of a critical value $\tilde{\gamma}$ such that when
$\gamma$ approaches $\tilde{\gamma}$ the ground state switches from $|\psi_0
\rangle$ to $|\psi_1 \rangle$, at least in the limit $N \rightarrow \infty$. In
fact, if we take $\tilde{\gamma} = C N^{\alpha} \zeta_1$, then in
Eq.~(\ref{eq:bound_ground}) and Eq.~(\ref{eq:bound_any}) we can set $\gamma =
(C+\epsilon) N^{\alpha} \zeta_1$, with $\epsilon >0$ and $\epsilon <0$
respectively, obtaining 
\begin{equation} \label{eq:bound_simply} 
1 - |\langle s
| \psi_{0,1} \rangle |^2 < \frac{C}{N^{1+\alpha}\epsilon^2}
\frac{\zeta_2}{\zeta_1^2}.  
\end{equation} 
Therefore, recalling that $\alpha
-1$, the condition $\zeta_2 = \mathcal{O}(\zeta_1^2)$, as $N \rightarrow
\infty$ is sufficient for $|\langle s | \psi_{0,1} \rangle |^2 \rightarrow 1$.
Otherwise stated, as $\gamma$ approaches $\tilde{\gamma}$ from above ($\epsilon >
0$) or from below ($\epsilon <0$), the overlap of the ground state with
$|\psi_0 \rangle$ and with $|\psi_1 \rangle$, respectively, is close to $1$.

\section{The Dual Sierpinski gasket}\label{appe2} The Laplacian spectrum
$\mathcal{E}(k)$ for the DSG is exactly known
\cite{cosenza,jurjiu,ABM2008} and we can therefore calculate exactly the
quantities $\zeta_1$ and $\zeta_2$, obtaining estimates for the critical
parameter $\tilde{\gamma}$.

At a given generation $g$, the spectrum includes the eigenvalue $3$ with
degeneracy $m(3,g)=(3^{g-1}+3)/2$, the eigenvalue $5$ with degeneracy
$m(5,g)=(3^{g-1}-1)/2$ and the eigenvalues $\lambda^+$ and $\lambda^-$ stemming
from the eigenvalue $\lambda$, belonging to previous generation and both
carrying degeneracy $m(\lambda,g-1)$. For each eigenvalue $\lambda$ the
next-generation eigenvalues $\lambda^{\pm}$ are defined according to
\begin{equation}\label{eq:iteration} 
\lambda^{\pm}= \frac{5 \pm \sqrt{25 - 4
\lambda}}{ 2} 
\end{equation} 
Now, it follows directly from
Eq.~(\ref{eq:iteration}) that any couple $1 / \lambda^+$, $1 / \lambda^-$ sum up
as 
\begin{equation}\label{eq:sum_reciprocal}
\frac{1}{\lambda^+}+\frac{1}{\lambda^-} =\frac{5}{\lambda}.  
\end{equation} 
and
applying this result iteratively to all the couples making up the spectrum we
get 
\begin{eqnarray}\label{eq:xi1_DSG} 
\zeta_1 & \equiv & \sum_{k \neq 0}
\frac{1}{\mathcal{E}(k)} \nonumber \\ &=&\sum_{j=0}^{g-1} \frac{5^j}{3} \left(
\frac{3^{g-j-1} +3}{2} \right) + \sum_{j=0}^{g-1} \frac{5^j}{5} \left(
\frac{3^{g-j-1} -1}{2} \right) \nonumber \\ &=&\frac{1}{30} (-3 -4\times 3^g + 7
\times 5^g).  
\end{eqnarray}
As for $\zeta_2$ we can implement a similar iterative procedure, by noticing
that 
\begin{equation}\label{eq:sum2_reciprocal}
\frac{1}{(\lambda^+)^2}+\frac{1}{(\lambda^-)^2} =\frac{25-2 \lambda}{\lambda^2}.
\end{equation} 
Therefore $\zeta_2$ is made up of two terms stemming from the
first and second order contributions, which are, respectively, 
\begin{eqnarray}
\nonumber & & \sum_{j=0}^{g-1} \frac{5^{j-1}}{3}  \frac{(5^{j} -1)}{2}
\frac{3^{g-j-1} + 3}{2} \\ &+& \sum_{j=0}^{g-1} \frac{5^{j-1}}{5} \frac{ (5^{j}
-1)}{2}  \frac{3^{g-j-1} -1}{2} \nonumber \\ &=& \frac{55 + 80 \times 3^g -154
\times 5^g +19 \times 5^{2g}}{6600} 
\end{eqnarray} 
and
\begin{eqnarray}\label{eq:xi2_DSG} 
\nonumber \sum_{j=0}^{g-1} \frac{25^j}{3^2}
\frac{3^{g-j-1} + 3}{2} + \sum_{j=0}^{g-1} \frac{25^j}{5^2} \frac{3^{g-j-1}
-1}{2} \\ = \frac{-121 -68 \times 3^g +189 \times 5^{2g}}{19800}.
\end{eqnarray} 
Subtracting Eq.~(\ref{eq:xi2_DSG}) and Eq.~(\ref{eq:xi1_DSG}) we
finally get 
\begin{equation}\label{eq:xi_DSG} 
\zeta_2 \equiv \sum_{k \neq 0}
\frac{1}{\left [ \mathcal{E}(k) \right ]^2} = \frac{1}{900} (-13 -14\times 3^g +
21 \times 5^g + 6 \times 25^g).  
\end{equation}

Now, recalling that $N=3^g$ and that the spectral dimension of the DSG is
$\tilde{d}=2 \ln 3 / \ln 5$, we can write $5^g = N^{2/\tilde{d}}$ and obtain
expressions for $\zeta_1$ and $\zeta_2$ as a function of just the volume $N$ and
the spectral dimension $\tilde{d}$ of the substrate:
\begin{equation}\label{eq:xi1_N} 
\zeta_1 = \frac{1}{30} \left ( -3 -4 N + 7
N^{2/\tilde{d}} \right ) \sim \frac{7}{30} N^{2/\tilde{d}}, 
\end{equation}
\begin{equation}\label{eq:xi2_N} 
\zeta_2 = \frac{1}{900} \left ( -13 -14 N +21
N^{2/\tilde{d}} + 6 N^{4/\tilde{d}}\right ) \sim \frac{1}{150} N^{4/\tilde{d}},
\end{equation} 
where the asymptotic expressions hold for large $N$.

We notice that as $N \rightarrow \infty$, $\zeta_1$ and $\zeta_2$ appearing in
Eqs.~(\ref{eq:xi1_N}) and (\ref{eq:xi2_N}) satisfy the condition $\zeta_2 =
\mathcal{O}(\zeta_1^2)$ found in Appendix \ref{appe1}. More precisely, $\zeta_2
\sim \zeta_1^2$ and this is sufficient for $1-|\langle s | \psi_0 \rangle|^2$
and $1-|\langle s | \psi_1 \rangle|^2$ to converge to zero as $N^{-\alpha}$. In
particular, from Eq.~(\ref{eq:xi1_N}) and Eqs.~(\ref{eq:bound_ground}) and
(\ref{eq:bound_any}), the state $| s \rangle$ is expected to switch from the
ground to the first excited state at 
\begin{equation}\label{eq:gammacrit}
\tilde{\gamma} \approx C N^{2/\tilde{d}+\alpha}.  
\end{equation}

As explained in Sec.~\ref{sec:phase_transition}, the expressions found here for
$\zeta_1$ and $\zeta_2$ are consistent with $\zeta_1=\sum_k 1/[\mathcal{E}(k)]^j
\sim N^{2j/d}$ obtained in \cite{childs} for lattices. Therefore, and as
suggested by the numerical data discussed in Sec.~\ref{sec:numerics}, it is
plausible that Eqs.~(\ref{eq:xi1_N}), (\ref{eq:xi2_N}) and, above all, the
expression for the critical parameter $\tilde{\gamma}$ in
Eq.~(\ref{eq:gammacrit}) can be extended to arbitrary structures of spectral
dimension $\tilde{d}<2$.

\section{The complete graph}\label{appe3} Let us start from the definition of
success probability given in Eq.~(\ref{eq:success_prob}).
Now, the propagator $\mathbf{U} \equiv \exp(- i \mathbf{H} t)$, can be
calculated as 
\begin{equation}
U_{k,j} =
\sum_{l=1}^{\infty} \frac{(-i t)^l}{l!} (\mathbf{H}^l)_{k,j}.  
\end{equation}
and the success probability can be rewritten as
\begin{equation}\label{eq:success_prob_1} 
\pi_{w,s}^{\gamma}(t) = \frac{1}{N}
\left| \sum_{k=1}^{N} U_{k,w} \right|^2.  
\end{equation}
Hence, in order to calculate the success probability $\pi_{w,s}^{\gamma}(t)$ we
need to find the elements corresponding to the $w$-th column of the $l$-th power
of the Hamiltonian. For complete graphs, $\mathbf{H}^l$ displays a high degree
of symmetry which allows its exact calculation (see for example \cite{farhi1}
for a similar calculation where the Hamiltonian is provided by the adjacency
matrix). Without loss of generality we can fix $w=N$ so that the Hamiltonian
$\mathbf{H}$ is 
\begin{equation}\label{eq:hamiltonian_mat} 
H=\gamma \left(
\begin{array}{ccccc} N-1 & -1 & -1 & \ldots & -1 \\ -1 & N-1 & -1 & \ldots & -1
\\ -1 & -1 & N-1 & \ldots & -1 \\ \vdots & \ddots & & & \vdots\\ -1 & -1 & -1 &
\ldots & N-1-1/\gamma \end{array} \right) 
\end{equation} 
and it is easy to see
that, regardless of $l$, $(\mathbf{H}^l)_{i,N}=(\mathbf{H}^l)_{j,N}$, with
$i,j<N$. Therefore, Eq.~(\ref{eq:success_prob_1}) can be rewritten as
\begin{equation}\label{eq:success_prob_2} 
\pi_{w,s}^{\gamma}(t) = \frac{1}{N}
\left| (N-1)U_{1,N} +  U_{N,N} \right|^2.  
\end{equation} 
Now, our task is to
calculate $U_{1,N}$ and $U_{N,N}$ for which we need the entries $(1,N)$ and
$(N,N)$ of the $l$-th power of the Hamiltonian; for better readability we set
$a_{1N}(l) \equiv (\mathbf{H}^{l})_{1,N}$ and $a_{NN}(l) \equiv
(\mathbf{H}^{l})_{N,N}$. Thus, from Eq.~(\ref{eq:hamiltonian_mat}) we derive the
following recursive relations: 
\begin{equation}\label{eq:recursive_1}
a_{1N}(l+1) = a_{1N}(l)-a_{NN}(l), 
\end{equation} 
and
\begin{equation}\label{eq:recursive_2} 
a_{NN}(l+1) =
-(N-1)a_{1N}(l)+(N-1-\gamma^{-1})a_{NN}(l).  
\end{equation} 
From their
combination we get 
\begin{equation}\label{eq:recursive_3} 
a_{1N}(l+2) =
(N-\gamma^{-1}) a_{1N}(l+1)+ \gamma^{-1}a_{1N}(l).  
\end{equation} 
whose
solution is 
\begin{equation}\label{eq:a1va} 
a_{1N}(l) = \frac{\gamma}{2^l}
\frac{1}{B} \left[ (A+B)^l - (A-B)^l \right], 
\end{equation} 
with 
$A=N \gamma
-1$ and $B=\sqrt{A^2 + 4 \gamma}$.  From Eq.~(\ref{eq:recursive_1}) and
Eq.~(\ref{eq:a1va}), $a_{NN}(l)$ is also explicitly defined:
\begin{equation}
a_{NN}(l) = \frac{\gamma}{2^l} \frac{1}{B} \left[
(A+B)^l - (A-B)^l \right], 
\end{equation} 
Now, according to
Eq.~(\ref{eq:success_prob_1}), we can calculate $U_{1,N}$ as
\begin{eqnarray}\label{eq:a1v} 
\nonumber U_{1,N} &=& \sum_{l=0}^{\infty}
\frac{(-i t \gamma)^l}{l!}  \frac{\gamma}{2^l} \frac{1}{B} \left[ (A+B)^l -
(A-B)^l \right] \\ \nonumber &=&\frac{\gamma}{B} \left \{ \exp{\left[-it(A+B)/2
\right]} -\exp{\left[-it(A-B)/2 \right]}  \right\}\\ &=&\frac{\gamma}{B} \exp{-i
t A/2} (-2 i \sin{\frac{tB}{2}}).  
\end{eqnarray} 
Analogous calculations lead to
\begin{eqnarray}\label{eq:avv} 
\nonumber U_{N,N} &= & \frac{\exp{(-it/2A)}}{B}
\\ & \times & \left ( -2i \gamma \sin{\frac{tB}{2}} + i A \sin{\frac{tB}{2}} - B
\cos{\frac{tB}{2}} \right).  
\end{eqnarray}

Hence, by inserting Eqs.~(\ref{eq:a1v}) and (\ref{eq:avv}) into
Eq.~(\ref{eq:success_prob_2}), and performing the summation
\begin{eqnarray}\label{eq:success_prob_3} 
\nonumber \pi_{w,s}^{\gamma}(t) &=&
\frac{1}{N} [ 1 + \frac{4 \gamma (N-1)}{1+4 \gamma - 2 N \gamma + N^2 \gamma^2 }
\\ & \times & \sin^2{(t(\sqrt{1+4 \gamma - 2 N \gamma + N^2 \gamma^2})/2)} ],
\end{eqnarray} 
which provides the exact success probability for the complete
graph of size $N$ as a function of $\gamma$ and of $t$.  We can notice that
$\pi_{w,s}^{\gamma}(t)$ is strictly larger than zero and that it can be equal to
$1$ when both $t=\pi (2k+1)/\sqrt{1+4 \gamma - 2 N \gamma + \gamma^2 N^2}$, with
$k \in \mathbb{Z}$ and $4 \gamma / (1+4 \gamma - 2 N \gamma + N^2 \gamma^2) = 1$
are satisfied. The latter condition holds for $\gamma N =1$, just consistent
with \cite{childs}: there it is found that for $\gamma N =1$ the walk rotates
the state from $|s\rangle$ to $|w\rangle$ and that the gap between the
corresponding energies is smallest. For $\gamma N =1$ the first condition reads
$t=\sqrt{N}\pi(k+1/2)$.  Hence, for the exact $\gamma$ and at the right time the
success probability is unitary, but for larger volumes the right times get
sparser. We also notice that when $N$ is large and $N \gamma \sim 1$,
Eq.~(\ref{eq:success_prob_3}) can be simply rewritten as
\begin{equation}\label{eq:success_prob_4} 
\pi_{w,s}^{\gamma}(\bar{t}) \approx
\sin^2{(t/\sqrt{N})} .  
\end{equation}
From Eq.~(\ref{eq:success_prob_3}) we deduce that for a given time $\bar{t}$,
$\pi_{w,s}^{\gamma}(t)$ decreases as $|\gamma N -1|$ gets larger.


\end{document}